# Inverse-Designed Non-Hermitian Hollow Nanowire Cavity for Generating Optical Orbital Angular Momentum

XUEN ZHEN LIM,[1,†] MASATO TAKIGUCHI,[1,2,†,*] RYUJI KURUMA,[3] HISASHI SUMIKURA[1,2], AND MASAYA NOTOMI[1,2,3]

[1]*Basic Research Laboratories, NTT, Inc., 3-1 Morinosato Wakamiya, Atsugi, Kanagawa 243–0198, Japan*
[2] *NTT Nanophotonics Center, NTT, Inc., 3-1 Morinosato Wakamiya, Atsugi, Kanagawa 243–0198, Japan*
[3]*Department of Physics, Institute of Science Tokyo, 2-12-1 Ookayama, Meguro-ku, Tokyo 152-8550, Japan*
[†]*These authors contributed equally to this work.*
**masato.takiguchi@ntt.com*

Abstract: We designed a gallium nitride hexagonal hollow nanowire whispering gallery mode cavity that generates an |m|=6 topological light with orbital angular momentum (OAM). By exploiting the unidirectional propagation that emerges in the vicinity of singularities in non-Hermitian systems, this mechanism enables the generation of OAM. OAM is generated by breaking the cross-sectional mirror symmetry of the nanowire, which creates a non-Hermitian system. This is achieved by replacing the central airhole of the hollow nanowire with a cluster of 6 overlapping circular air holes with rotational offset relative to the hexagonal cross-sectional profile of the nanowire. The design parameters were then further optimized in Finite Element Method using an inverse design method to maximize the normalized OAM order $|l|$. We were able to realize of a cavity mode with $|l|$ = 5.7, a mode purity of about 97%, and a Q-factor of ~250. This marks the OAM generating active photonic device design falling within a sub-micron footprint, with additional novelties of being single component and materialistically homogeneous.

## 1. Introduction

Orbital angular momentum (OAM)[1][2] has increasingly become a topic of interest in the field of photonics for its potential in communications, information processing, as well as laser technology. An impactful example of OAM is the introduction of a new degree of freedom to information multiplexing, whereby data bandwidth can be dramatically increased[3][4]. This can be achieved because information from different OAM modes is able to occupy the same frequency, polarization, and spatial path without interfering with each other as they form an independent basis set. Similarly, the benefit of OAM modes spanning an infinite discrete set of orthogonal states also extends to the field of quantum information[5], allowing for superposition of multiple OAM states (qudits) as opposed to the conventional two states (qubits) achieved through horizontal and vertical polarization. Lately, OAM has also been shown to be capable of introducing torque to optical tweezers[6], acting as an optical analog of a spanner or wrench, which allows for finely controlled physical manipulation and structural modification of microscopic objects[7]. Additionally, in the field of microscopic sensing and imaging, scientists were able to leverage the properties of OAM light and overcome the Rayleigh diffraction limit to improve spatial resolution[8].

OAM is generated through the implementation of a non-Hermitian system, which can be achieved and maximized by strategically breaking the mirror symmetry of the system. OAM generation in photonics system can be traced as far back as to the utilization of Computer-Generated Holograms[9], spiral phase plates[10], and spatial light modulators[11]. In the last decade, the implicated benefits of integrating OAM into photonic chips have inspired the creation of various miniaturized OAM photonic devices (especially for lasers[12]), such as iterations of micro-ring resonators[13][14][15][16] and meta surfaces[17][18]. These micro-scale systems often come with complicated structures, which may involve the coupling of multiple individual components[15], intricate manipulation of component geometry[19], and/or the utilization of several materials with varying refractive indices[14]. It can be reasonably inferred that the complex nature of these designs limits the minimum possible device footprint to the micrometer level, as further sub-micron downscaling of the same systems would require fabrication process at precision levels yet to be demonstrated.

As of late, the possibility of generating nanoscale OAM through a novel method of coupling a hollow standing gallium nitride (GaN) nanowire and a carefully positioned gold standing nanowire had also been realized in Finite Element Method (FEM)[20]. A viable fabrication process for the main component to this system, the hollow standing GaN nanowire, has already been demonstrated recently and shown to be capable of generating topological modes[21]. Despite that, this OAM generation approach still falls short of becoming practically feasible as there has yet to be a method to fabricate a positionally precise, free-standing gold nanowire, which is the crucial component responsible for introducing non-Hermitian perturbation to the system. Another recent example demonstrates a fabricable sub-micron OAM generation device using $TiO_2$ meta-quadrumers, whereby $TiO_2$ nanopillars with varying radii are placed in a 2x2 array configuration[22]. Like many existing OAM generation systems, this device still relies on external light source to operate, as $TiO_2$ is a passive material, as opposed to GaN, which supports lasing.

In this study, we demonstrate, through FEM simulations, an OAM-generating nanoscale device that can be realistically fabricated using existing techniques. Figure 1 illustrates a conceptual schematic of the hollow nanowire OAM light source incorporating a non-Hermitian perturbation. In hollow hexagonal nanowires with a circular inner boundary preserving mirror symmetry, there are only standing-wave resonant states consisting of pairs of even and odd modes, which do not exhibit chirality. In order to achieve OAM, we have to introduce asymmetric loss to this system. Here, we introduce a periodic perturbation for the inner boundary with a finite off-set angle to the original hexagon, which breaks the mirror symmetry of the hexagonal cross-section and produces appropriate asymmetric radiation loss, thereby inducing non-Hermicity onto the system's Hamiltonian. The maximum chirality can be obtained by adjusting the off-set angle.

This design uses the same hollow standing gallium nitride nanowire mentioned above as the main component, with a modification towards the center of the cross section, whereby the circular airhole is replaced with a cluster of 6 overlapping, radially offset airholes. The radius of the nanowire, along with the radius, radial offset, and rotational offset of the airholes were then optimized using a novel inverse design method, Fuzzy Self-Tuning Particle Swarm Optimization. The rotated central airhole cluster, relative to the outer hexagonal shape of the nanowire, induces an asymmetric in-plane radiation loss, resulting in a non-Hermitian perturbation on the system. We managed to realize OAM in a transverse magnetic cavity mode with an order of $|l|$ = 5.7, a mode purity of about 97%, and a Q-factor of ~250.

Such GaN-based structures can be realized using established fabrication techniques, including sublimation [21][23][24] and dry etching processes [25][26][27]. Unlike a typical ring resonator, where both clockwise and counterclockwise modes coexist, this structure can selectively generate either positive or negative OAM purely through its design, without requiring connections to waveguides or other components. With the additional factors of being an active system, single component, and materialistically homogeneous, this device design possesses an unprecedented combination of properties that opens new potential for ultra-compact OAM device integration.

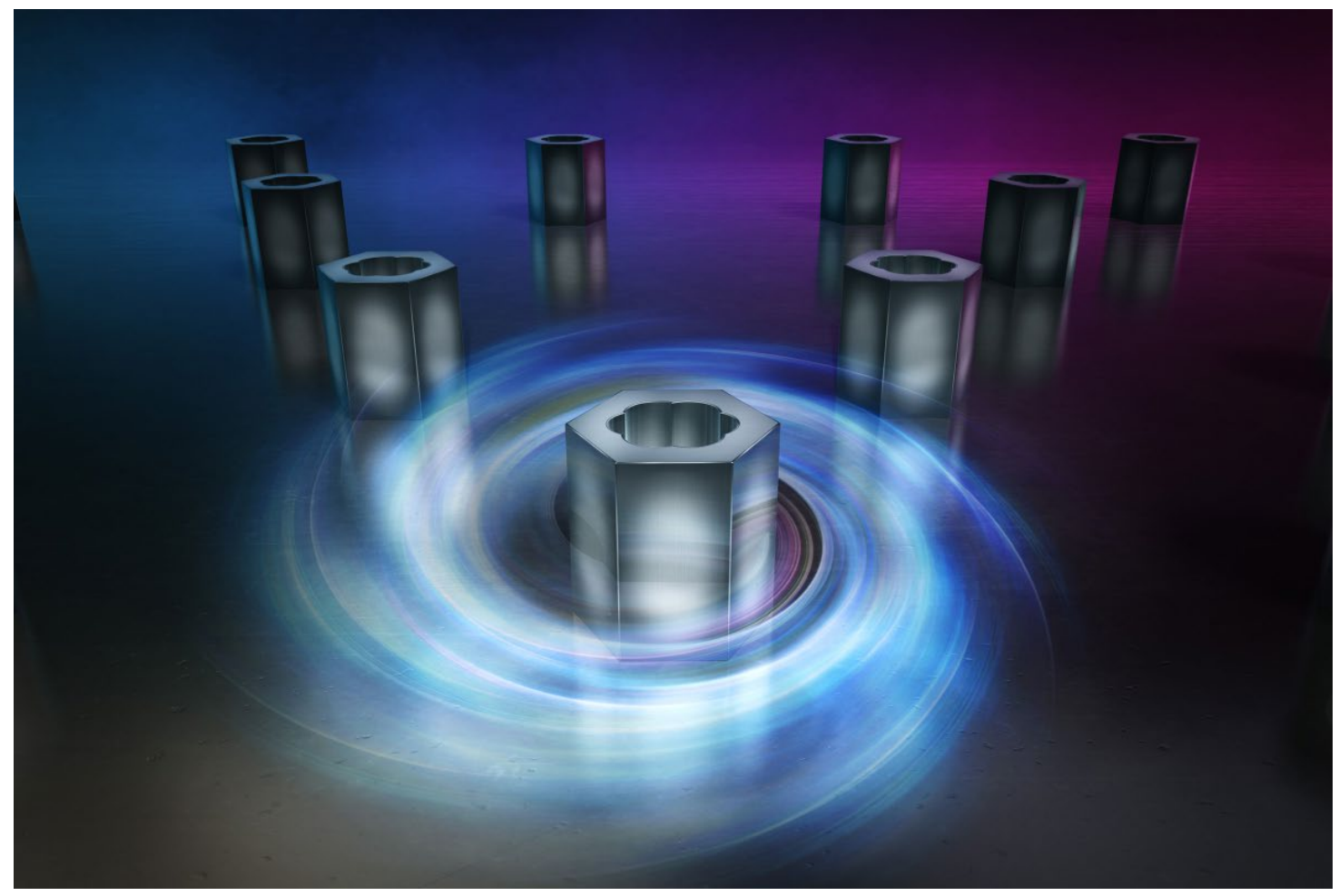

Fig. 1. Schematic hollow nanowire cavity

## 2. Inverse Design

Various inverse-design methods have been proposed and proven highly effective for the optimization of photonic devices[28] and nanowire cavities[29]. In this study, the parameters of the hollow GaN nanowire design were optimized in a Python program with Fuzzy Self-Tuned Particle Swarm Optimization (FST-PSO)[30], a modified version of the Particle Swarm Optimization (PSO) algorithm[31] (Supplemental 1). Our program uses the 'fst-pso' library to perform the calculations for parameter tuning and integrates the 'MPh' library to interact with the COMSOL simulation to execute calculations for the respective parameter sets and extract calculated results. The 'multiprocessing' library was used for thread assignments to have parallel COMSOL sessions running concurrently. The swarm was initiated with 12 particles, with each particle running a single COMSOL session. 40 CPU cores were allocated to be shared across the 12 instances, whereby the core usage would be dynamically distributed depending on the remaining ongoing instances.

The optimized parameters include the radius of the hexagonal nanowire ($r_{\text{hex}}$), the radius and radial offset of the central airhole ($r_{\text{hole}}$ and $R$), and the height of the nanowire ($h$) (Fig. 2(a)). 5 additional airhole were then generated by rotating the central airhole over the center of the nanowire by an interval of 60° to create an airhole cluster (Fig. 2(b)). Finally, a rotational offset was applied to the airhole cluster ($\Delta\theta$) (Fig. 2 (c)). The fitness, or objective, used for the optimization was set as the OAM order (topological charge, azimuthal mode order) $|l|$ calculated within the GaN medium of the nanowire. To quantitatively evaluate the angular momentum characteristics of the electromagnetic field in a cylindrically symmetric coordinate system $(r, \phi, z)$, we consider the normalized z-component of the angular momentum density, $J_z/W$, where $J_z$ represents the angular momentum flux and $W$ denotes the total energy density. The normalized angular momentum density[32] is given by:

$$l = \frac{J_z}{W} = \frac{\text{Im}\int\int \mathbf{E}^*(r,\phi,z)\frac{\delta}{\delta\phi}\mathbf{E}(r,\phi,z)rdrd\phi}{\int\int \mathbf{E}^*(r,\phi,z)\mathbf{E}(r,\phi,z)rdrd\phi}. \qquad (1)$$

To analyze the OAM content of an optical field in cylindrical coordinates, we compute the azimuthal Fourier spectrum of the complex electric field at a fixed longitudinal plane $z = z_0$. This decomposition allows us to extract the relative weight of each m, corresponding to distinct helical phase modes $e^{im\phi}$. The OAM spectrum $R_l$ [33][34]and Fourier coefficient for each m are defined as:

$$R_m = \frac{|F_m|^2}{\sum_{-\infty}^{\infty}|F_m|^2} \qquad (2)$$

and

$$F_m = \int_0^{2\pi}\int_0^{\infty}\mathbf{E}(r,\phi,z_0)\, e^{-im\phi}rdrd\phi\,. \qquad (3)$$

We performed the optimization in 2D FEM, due to the relatively lower computational demand, to explore and conceive the initial cross-sectional design that produces OAM. The objective is limited to only extracting the OAM value of modes with Q-factor of at least 100 and confinement factor Γ of at least 50%, otherwise the OAM order is set to 0. Once we identified a potential design and the topological order supported (in our case, |m|=6), we recreated the cross-section design in 3D FEM, so that we can further optimize the nanowire design in a more realistic condition. The cross section was extruded uniformly in the z-axis to create the GaN nanowire and was positioned to be standing on top of a sapphire substrate surrounded by a cylindrical scattering boundary, and the eigenfrequency solver under COMSOL. The 3D optimization approach was similar to that of 2D, with the height of nanowire set as an additional optimization parameter. The same objective extraction logic from 2D optimization is also applied to 3D but with an extra limitation that the mode must simultaneously have a spectral purity of at least 80% for |m|=6 ($F_{+6} + F_{-6} \geq 0.8$) and must possess the highest Q-factor among the neighboring modes.

For both 2D and 3D cases, 'Electromagnetic Waves, Frequency Domain' physics module was used for the calculation. We evaluated OAM density by calculating Eq. (1) over the nanowire's cross section, using integral probe in the 'Domain Probe' feature. As for determining the mode purity and Fourier spectrum, we utilized the 'Integration' feature over the nanowire cross section to define Eq. (2) as a function of azimuthal order $m$. We derive mode purity for the hexagonal nanowire by defining m=-6, 6. The denominator in Eq. (2) acts as the normalization function that examines the Fourier coefficient across infinite azimuthal orders, but for this implementation it is constrained to azimuthal orders ranging from −20 to 20. The swarm optimization was set to run until 10 epochs without a new global best fitness being reached. The inclusive boundaries used for parameters are as shown in the Table 1 below.

| Parameter | Symbol | Unit | Lower bound | Upper bound |
| --- | --- | --- | --- | --- |
| Radius of the central airhole | $r_{\mathrm{hole}}$ | m | $5 \times 10^{-8}$ | $11 \times 10^{-8}$ |
| Radial offset of the central airhole | $R$ | m | $1 \times 10^{-8}$ | $9 \times 10^{-8}$ |
| Rotational offset | $\Delta\theta$ | ° | 10 | 20 |
| Radius of the hexagonal nanowire | $r_{\mathrm{hex}}$ | m | $2 \times 10^{-7}$ | $4 \times 10^{-7}$ |
| Height | $h$ | m | $1 \times 10^{-7}$ | $5 \times 10^{-7}$ |

Table 1. Inclusive boundaries set for each parameter. Note that nanowire height, $h$, only applies to optimization in 3D.

## 3. 2D Simulation

To estimate the mode profile and OAM, we used FEM. As shown in Fig. 2, our structure based on GaN is designed by replacing the central airhole of a hollow nanowire with 6 circular airholes overlapping each other. The target wavelength of the GaN-based cavity is ~400 nm (no non-negligible absorption losses). Upon optimization to maximize OAM, the values for $r_{\mathrm{hole}}$, $R$, $\Delta\theta$, and $r_{\mathrm{hex}}$ are 93 nm, 41 nm, 12°, and 216 nm respectively. Fig 3(a) shows the electric profile of the optimized hollow nanowire cavity (TM mode). As shown in Fig3 (a), the electric field ($E_z$) has a swirling tail. This suggests the characteristics of an optical vortex. In addition, clockwise and counterclockwise vortices were confirmed in the antisymmetric structure (Fig3 (d)). To directly verify this, the phase distribution is also plotted in Fig. 3(b) and (e). This clearly confirms the generation of OAM. After simulating the electric field distribution, the OAM order and mode purity was calculated using equations (1) and (2), as described in Section 2. The Fourier spectrum plot, shown in Figure 3(c) and (f), is calculated using equation (2) in a similar manner as mode purity, but with the numerator iterated through azimuthal orders ranging from −20 to 20 to represent each point across the graph. to represent each point across the graph. In addition, the time-averaged electric field intensity (Supplemental 1) shows 'smeared' anti-nodes relative to what is usually observed of standing wave WGMs, strongly indicating rotational movement in cavity. We also calculated and plotted the time-averaged Poynting vectors and magnitude distribution (Supplemental 1), showing high concentration of curl in cavity, especially when compared against a regular GaN nanowire with circular hollow core, further solidifying the presence of OAM. Through this optimization, the structure is able to support an |m|=6 WGM with normalized OAM density of $|l| = 5.7$. As shown in the OAM spectrum, the modes with opposite signs of OAM exhibits high spectral purity, exceeding 0.97.

Figure 4 shows the normalized OAM density values as a function of the offset angle $\Delta\theta$. From this figure, it is evident that the OAM reaches its maximum when the offset angle is set to 12°. As shown in the figure, two distinct modes possess OAM (Each frequency is approximately in the vicinity of $8.11\times10^{14}$ Hz). This indicates that the degeneracy present in a simple ring-shaped structure is lifted due to the polygonal shape and the inner hole configuration. As shown in Supplemental 1, when the inner airhole cluster is rotated, the difference between the inner and outer diameters of the nanowire clearly reveals a structural mirror-asymmetry.

To verify that this is indeed a non-Hermitian system, we note that non-hermicity of a Hamiltonian is characterized by the approach towards an exceptional point (EP): the coalescence or near coalescence of eigenvalue and eigenstate pair. EP vicinity as observed in eigenvalues can be determined through the crossing of either the real or imaginary component of eigenvalue pair as we vary some variables which control the strength of non-Hermitian perturbation[35]. In Wiersig's[36], a framework that also achieves non-Hermitian perturbation through angular offset between two types of perturbation sources, eigenfrequency crossing is visualized by sweeping the values of angular offset. Here, we achieve that by performing parametric sweep across $\Delta\theta$ from 0° to 60° in FEM, where we find a level repulsion and width crossing at the parameters of OAM maxima ($\Delta\theta$=12° and 48°) (Fig. 4).

The eigenstate pairs of WGM-based systems can be expressed in either the traveling wave basis (CW/+m or CCW/-m) or standing wave basis (even and odd parity modes). And, like spin angular momentum, their relationship can be mapped onto a Poincare sphere[37] (Supplemental 1) . The Hamiltonian of a WGM-based cavity possess solution with relative phase-inverted eigenstate pair existing along the surface of a Poincare sphere, which arise as (quasi-)degenerate standing modes in FEM simulation. EP approach of eigenstate pairs in WGMs simply corresponds to increased collinearity between the solution pair on the WGM Poincare sphere (both eigenvectors approximately pointing towards the same pole), which can be computed through finding the electric field profile overlap between the eigenmode pair in FEM. Equation 4 is implemented using the 'Integration' and 'Global Variable Probe' function over the nanowire's cross section domain in COMSOL, with the electric field index set to correspond to the WGM pair found in each solution.

$$S_S = \frac{\left|\int \langle E_a | E_b \rangle dV\right|}{\sqrt{\int |E_a|^2\, dV * \int |E_b|^2\, dV}} \qquad (4)$$

Likewise, we observed maximized overlaps at 12° and 48° when we sweep across $\Delta\theta$. These observations confirm EP vicinity of the Hamiltonian for the GaN nanowire structure. Additionally, we also compare this model against Wiersig's framework and explore the usage of two mode approximation for predicting collinearity Supplemental 1.

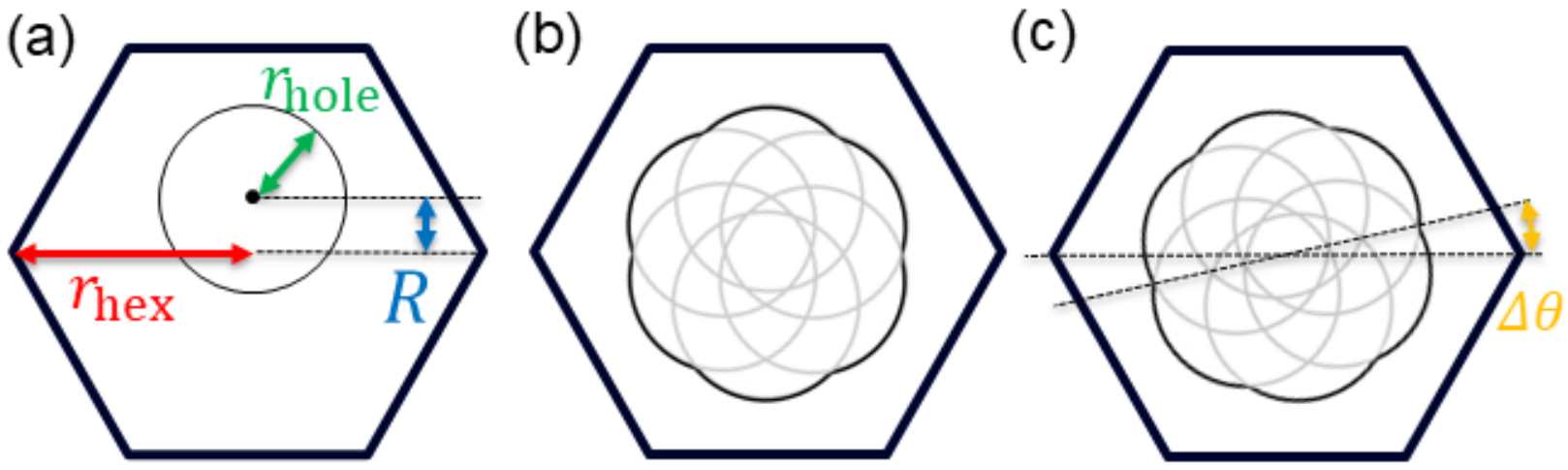

Fig. 2. (a) Top view of GaN nanowire, with radius of the hexagonal nanowire ($r_{\text{hex}}$), the radius and radial offset of the central airhole ($r_{\text{hole}}$ & $R$). (b) Depicts how the modified central airhole is rotated at 60 degrees interval, forming an airhole cluster from 6 overlapping circular airholes. (c) Rotational offset applied to airhole cluster ($\Delta\theta$), creating the finalized nanowire cross section design with asymmetric geometry.

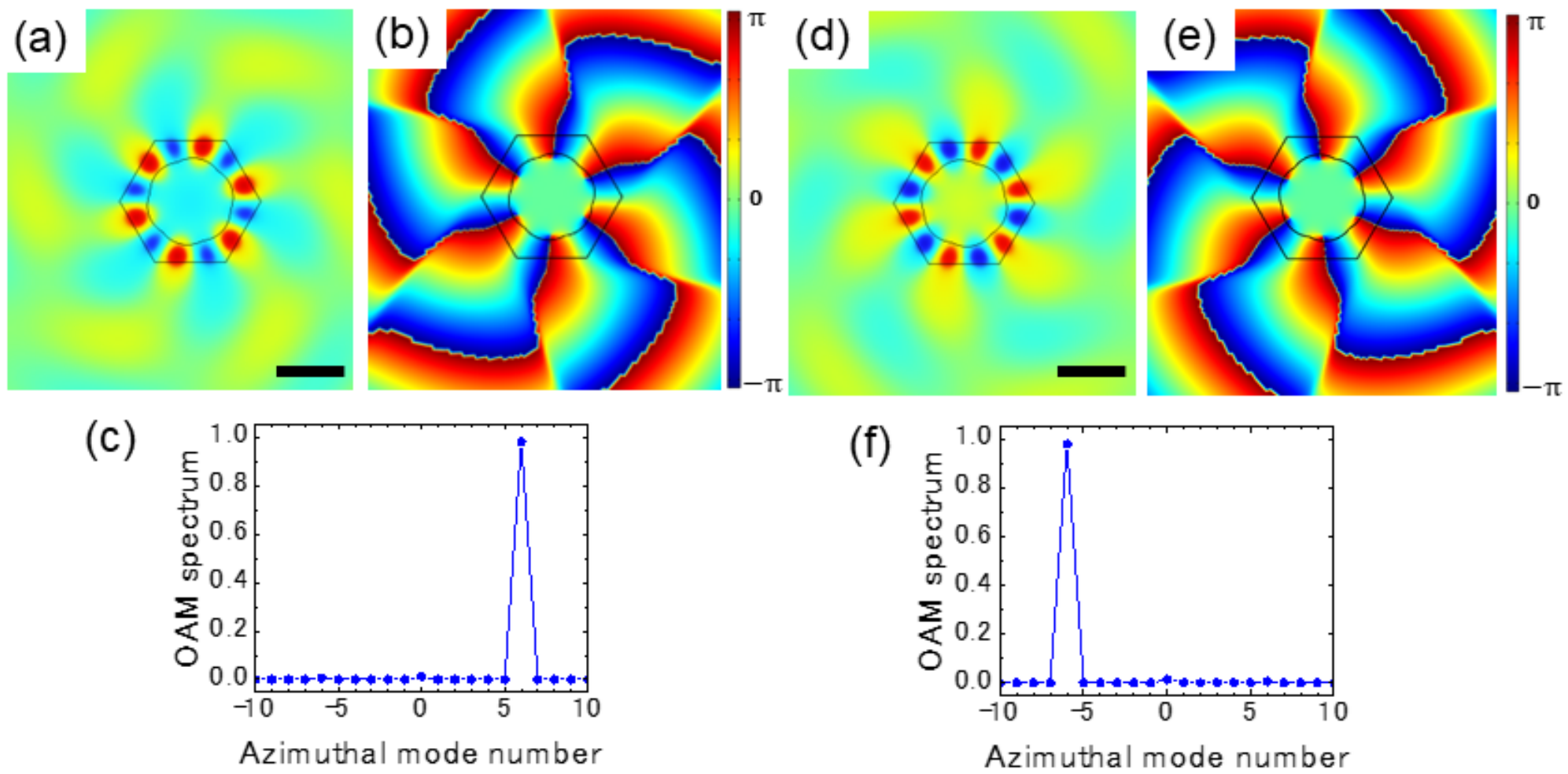

Fig. 3. (a) $E_z$distribution (scale bar: 200 nm). (b) Phase distribution (the color bar ranges from $-\pi$ to $\pi$). (c) OAM spectrum showing a peak at $l = 6$ with a purity of about 97%. (d) $E_z$distribution (scale bar: 200 nm). (e) Phase distribution (the color bar ranges from $-\pi$ to $\pi$). (f) OAM spectrum, where a peak appears at $l = -6$ when the cross-sectional geometry is mirrored.

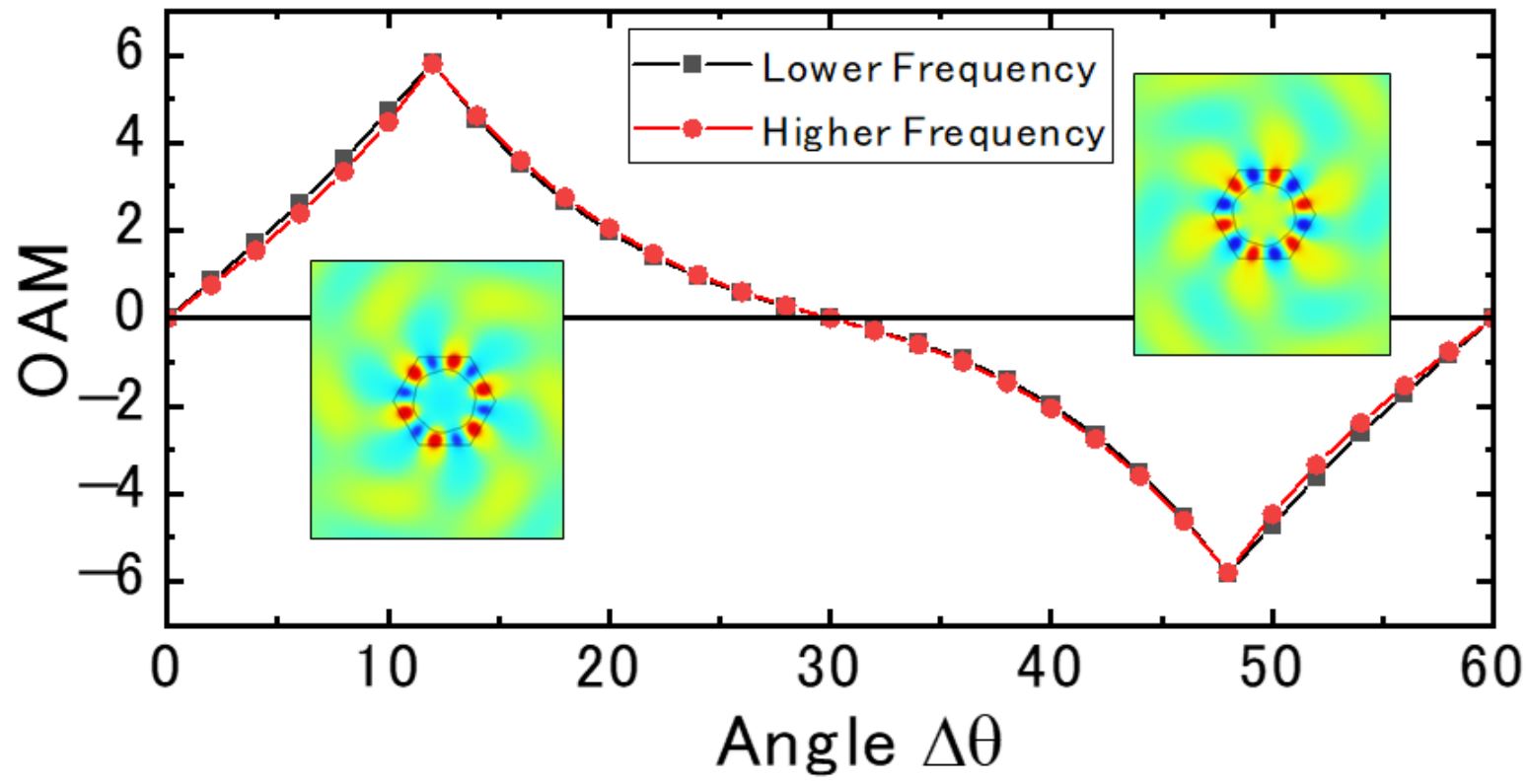

Fig. 4. Δθ dependence of OAM

## 4. Tolerance to Fabrication Errors

Next, we evaluate the robustness of the proposed design against fabrication imperfections. As shown in Fig 4, the OAM remains large even when the relative rotation angle between the hexagonal cross-section and the internal cluster of circular holes is varied by several degrees. Simulation results indicate that high OAM is maintained even under such angular deviations. Considering the typical fabrication tolerance of electron-beam lithography, this level of angular variation does not pose a significant issue. Another common source of fabrication error is deviation of the hole diameter from its designed value. Such deviations frequently occur in nanophotonic fabrication processes. For example, in photonic crystal fabrication, discrepancies between the designed pattern size and the transferred pattern in the resist or hard mask, as well as faster-than-expected lateral etching during the hole etching process, can result in variations in the final hole diameter. To evaluate this effect, we examined the change in OAM when the hole diameter of the optimized structure was varied. This was done by performing a parametric sweep in which the combined value of $r_{hole} + R$ was shifted by up to $\pm 5$ nm. As shown in Fig. 5 (a), for hole diameter variations of up to 5 nm, the modal purity remains above 85% (less than a 13% decrease from the peak value of ~97%), indicating that the OAM is largely unaffected by such dimensional deviations.

Surface roughness is another important factor that can significantly influence the performance of nanophotonic devices. In our simulation, surface roughness was introduced both on the outer surface of the nanowire and on the inner hole boundaries (Fig. 5(b)). The details of the computational model are provided in the Supplemental 1. We evaluated the mode purity over all combinations of amplitude $A$ from 0 to 5 nm and angular wavenumber $B$ from 10 to 50. The calculated mode purity is shown in Fig. 5(c). In practical semiconductor nanofabrication, surface roughness typically consists of very fine features, making the regime of large $A$ and $B$ particularly relevant. As can be seen from the figure, the OAM does not significantly degrade and remains above 90%. We observe that at points where defect angular wavenumber B equals to a multiple of the azimuthal order $m$ ($B = k \cdot m$), there are relatively greater dip in mode purity especially at lower order $k$. These cases correspond to when each side of the nanowire surface has identical imperfections. However, it is unlikely that perfectly matching fabrication roughness will arise without deliberation in the fabrication process. For the remaining angular wavenumbers ($B \neq k \cdot m$), we see little variation across the $B$-axis, with mainly $A$ having an inversed effect on mode purity. In practical fabrication, the surface roughness is typically irregular and in the order of a few nanometers. Therefore, even when considering realistic fabrication processes, the device exhibits sufficient robustness against fabrication imperfections, indicating that the proposed structure is experimentally feasible.

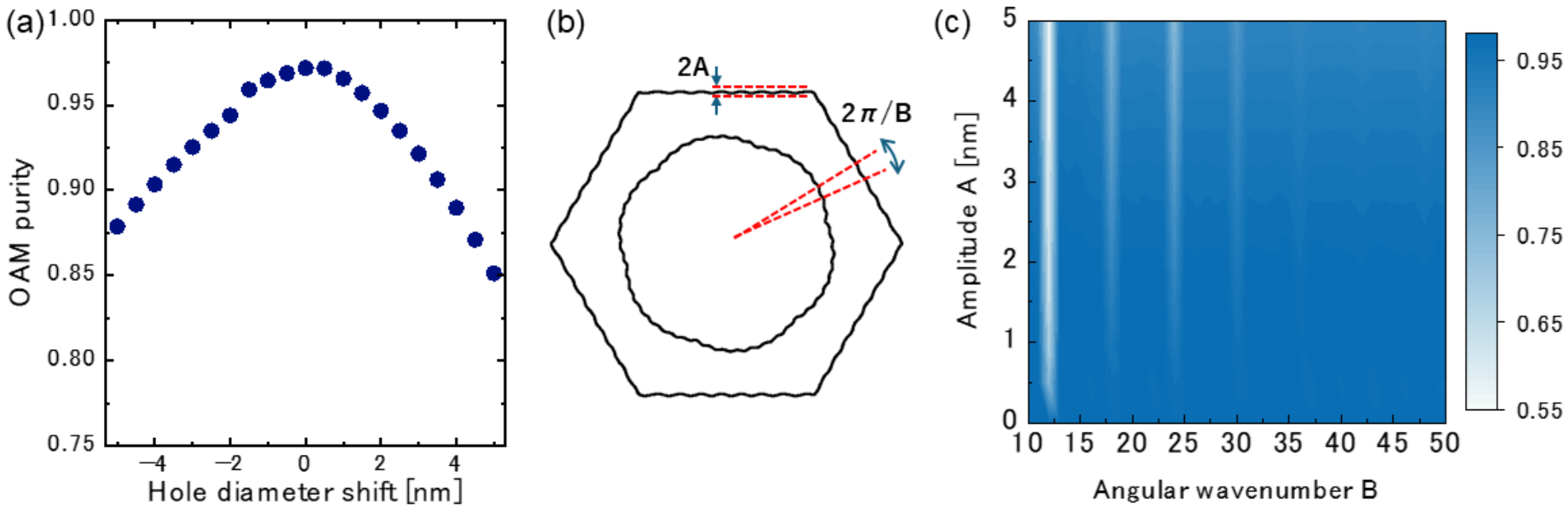

Fig. 5. (a) OAM purity dependence on diameter of airhole cluster (b) Simulation model (c) OAM mode purity dependence on defect amplitude $A$ and angular wavenumber $B$.

## 5. Polygonal Nanowire Cavities for OAM Control

Subsequently, we performed optimization for cavities with different N-sided polygon shapes, considering cases with N = 3 to 6. The results are shown in Fig. 6. We find that introducing $N$ overlapping air holes into an N-sided polygon structure most effectively enables the generation of OAM with topological charge |m|=N. However, as $N$ decreases, the induction of OAM becomes increasingly difficult. To address this, corner fillets were introduced as an additional optimization parameter, based on the hypothesis that the high radial discontinuity in polygons with smaller N may hinder efficient OAM generation. After including the corner radius in the optimization, we obtained cross-sectional designs with improved modal purity, with lower-N polygon structures requiring larger corner radii to achieve convergence. This trend is also observed in the hexagonal nanowires shown in Figures 3(a) and 6(d), where a slight improvement in OAM performance is evident in the structure with rounded corners, as indicated by the increased smoothness of the phase profile. These results demonstrate that the order of the generated OAM can be flexibly controlled through structural modification of the nanowire, providing versatility for a wide range of applications.

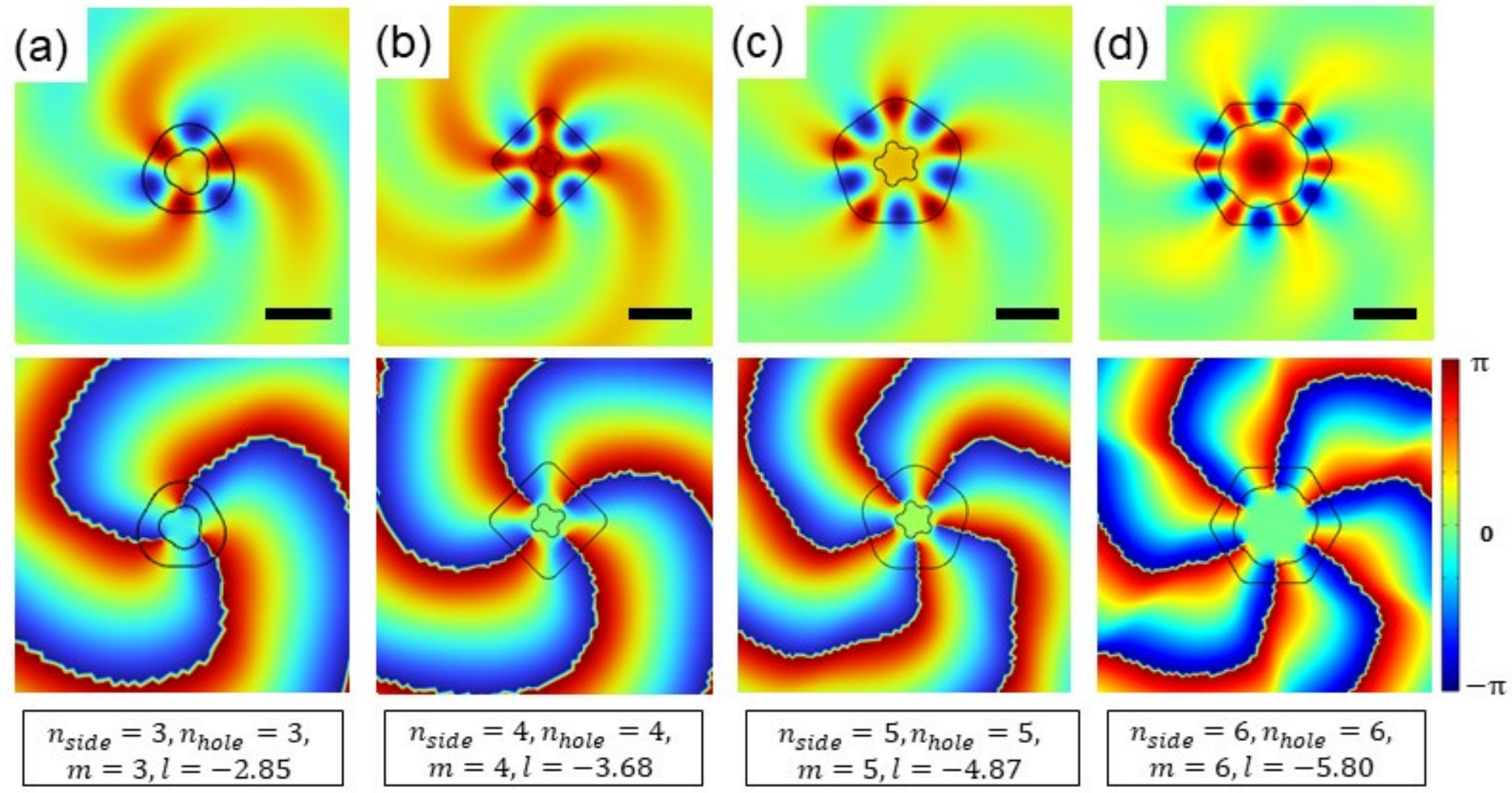

Fig. 6 (a)–(d) $E_z$distribution and phase distribution (color bar ranges from $-\pi$ to π; scale bar: 200 nm) for (a) $N = 3$, (b) $N = 4$, (c) $N = 5$, and (d) $N = 6$.

## 6. 3D Simulation

Finally, 3D optimization and calculation were performed to estimate the Q value. The final optimized values for $r_{\text{hole}}$, $R$, $\Delta\theta$, $r_{\text{hex}}$, and height are 80 nm, 52 nm, 12°, 229 nm, and 400nm respectively, with an eigenfrequency of 7.56 $\times 10^{14}$ Hz. This required an advanced meshing strategy tailored for the simulation of asymmetric OAM structures. Accurate modeling of orbital angular momentum (OAM) in three-dimensional photonic systems demands not only precise physical representation but also careful control of meshing. In this work, we developed a novel meshing technique that ensures high computational fidelity, even when geometric symmetry is partially or completely broken. Inaccurate meshing, especially under asymmetric conditions, can lead to spurious OAM features, potentially distorting physical interpretations.

To further optimize computational efficiency, we employed swept meshing in regions where geometry allowed it. The mesh structure used in the actual calculations is shown in the Supplemental 1. Compared to conventional tetrahedral meshing, swept meshing achieves equivalent numerical accuracy in geometries translationally invariant along an axis, while requiring only one-seventh the number of elements and half the degrees of freedom (DOF). This significantly decreased simulation time and memory consumption. We also incorporated boundary layer meshing near critical regions such as high-index contrast interfaces and scattering boundaries. By refining mesh fidelity in the direction of wave propagation and maintaining coarser meshing in distant regions, we preserved the accuracy of absorbing boundary conditions (e.g., SBC layers) without significantly compromising performance (For details of the mesh, refer to the Supplemental 1).

Figure 7 (a) and (b) show the simulation result. Here, the substrate is assumed to be sapphire, and the height of the nanowire is set to 400 nm. From this calculation, the mode showing WGM was identified. Despite the underlying system's asymmetry, the mesh configuration was carefully designed to be symmetrical, as it has a sensitive effect on the yielded value of OAM. We minimize the potential impact of meshing inconsistency by meshing the model in 24 equal sections, with each section possessing identical meshing elements As a result of this optimization, we realized a transverse magnetic cavity mode with $|l| = 5.7$, a modest Q of 250, as well as a mode purity (derived from the Fourier Spectrum equation in Eq. 2) of about 97% at the topological order of $|m|=6$ (OAM and spectral purity are measured at the middle cross section of the nanowire). Because the GaN-based materials used in this study do not have a particularly high refractive index, the resonator exhibits a relatively low Q factor; however, it may be possible to improve this in the future by using materials with a higher refractive index.

Furthermore, we examined the OAM outside the nanowire in the orthogonal direction. The calculated normalized OAM density $|l|$ at a plane 10 nm above the nanowire (i.e., in air) was 3.9, while at a plane 10 nm below the nanowire (i.e., in sapphire), it was 5. Since this resonator is based on z-invariant whispering-gallery modes (WGMs), radiation in the out-of-plane direction is inherently limited. To extract OAM light in the out-of-plane direction from this resonator, additional structures such as gratings would be required [15]. A cavity using similar principles to generate OAM for transverse electric (TE) WGMs could also be explored. Nonetheless, device applications that utilize this structure are conceivable, for instance, for particle trapping near the resonator in a liquid environment.

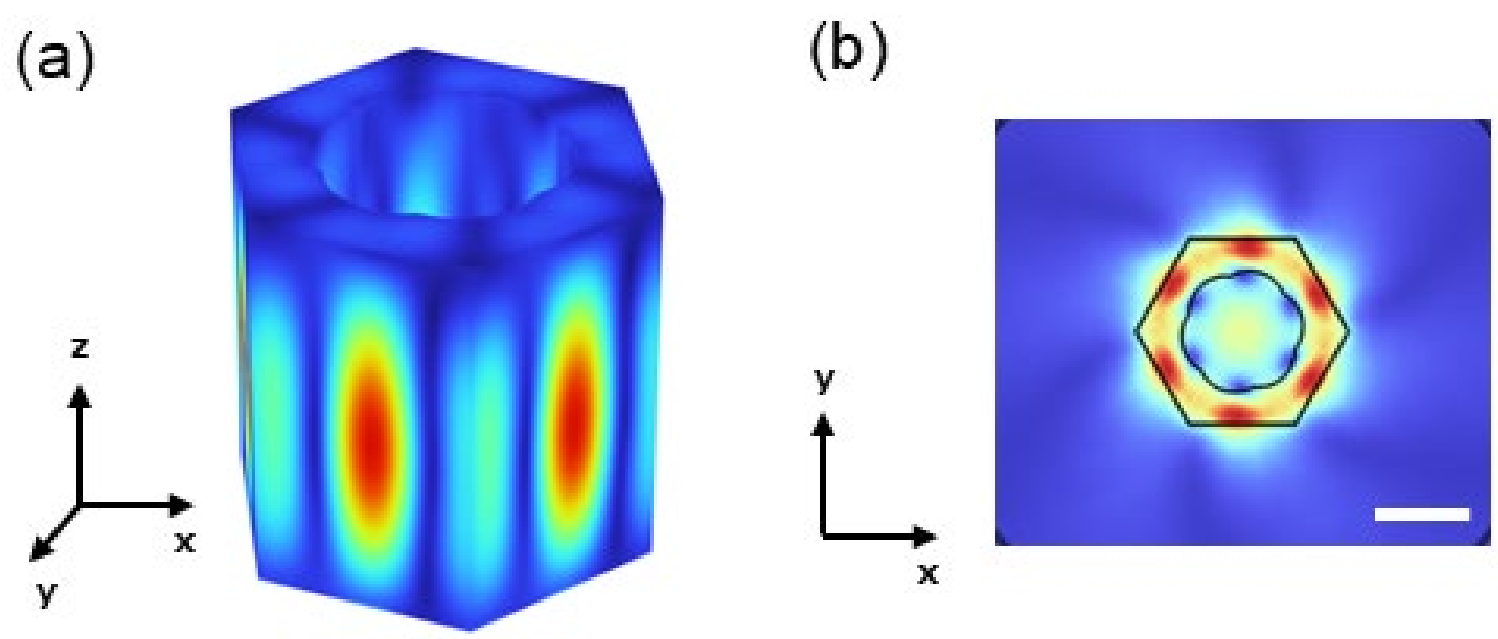

Fig. 7 (a) Cavity mode distribution of the hollow nanowire obtained from the 3D simulation. (b) Electric field distribution at the central cross-section of the nanowire (scale bar: 200 nm).

## 7. Conclusion and outlook

We demonstrated the selective extraction of a specific OAM by non-Hermitian perturbation, introducing scattering-induced losses. The device features a ring-shaped arrangement of six partially overlapping circular air holes, forming a gear-like pattern. This optimization enables the structure to support an $|m|=6$ WGM with a normalized OAM density of $|l| = 5.7$, achieving a high mode purity of 97% and a Q-factor about 250, consistent with a WGM of azimuthal order 6. The significance of this research lies not only in verifying the principles of non-Hermitian physics, but also in providing a novel, foundational design framework for ultra-compact OAM generating cavities by leveraging perturbation effects from deformations along both the inner and outer edges of a ring resonator-like cavity.

This cavity design can be useful for torque-enabled trapping of nano-sized particles[6], exceptional point-based sensing [38][39][40], as well as enabling high bandwidth channels for optical integrated computing and communications through OAM mode multiplexing [3][4]. As a more forward-looking application, it may also have potential relevance to radiation-pressure-driven systems such as solar sails in space[41] [42]. In this context, it may be possible to induce controlled rotational motion in nanoscale structures, thereby generating torque. In low-resistance environments such as outer space, this approach could potentially open new directions in nanomechanical systems.

In future work, it will also be necessary to investigate experimental measurement methods in more detail. The present resonator structure exhibits relatively strong in-plane scattering, while out-of-plane scattering is comparatively weak. In addition, the OAM observed in the far field is not necessarily identical to the near-field OAM (see the previous chapter). In general, the OAM emitted from nanolasers can be evaluated by using a lens with as high a numerical aperture (NA) as possible, self-interfering the far-field emission pattern, and analyzing the resulting interference fringes[43][44]. When constructing an interferometer is difficult, direct observation using a wavefront sensor is also possible[45][46]. Direct observation of the near field is not impossible either. For example, the azimuthal phase distribution can be experimentally accessed using phase-resolved near-field techniques such as Scanning Near-field Optical Microscopy (SNOM)[47]. As another possible approach, although it would provide only an indirect evaluation, one could also consider introducing a grating structure to enhance out-of-plane scattering of the emitted light [14].

In addition, as a possible direction for structural improvement, such OAM-generating nanowire resonators do not necessarily require the nanowire to be short. For example, by growing a long GaN nanowire and embedding multiple InGaN quantum well layers at its center, it is possible to confine light around the middle of the nanowire[48]. This scheme is also advantageous for nanowires, which can integrate an axial p–i–n structure, when implementing current-injection devices[49]. Moreover, the reciprocal nature of the emitted light can be exploited by coupling the nanowire to a waveguide, potentially enabling its use as an efficient light source. In this study, we discussed a single nanowire; however, it would also be interesting to investigate the collective behavior by fabricating an array structure of hollow nanowires. By utilizing metasurfaces or photonic bands, it may become possible to achieve even richer functionalities and characteristics. Thus, the present study is not merely a proof-of-concept using a toy model, but also demonstrates a promising approach with potential for future practical applications.

### Funding

This work has been supported by the JSPS KAKENHI Grant Number 15H05735, 23K26581 and 26K01425.

### Acknowledgments

We would like to express our gratitude to Peter Heidt, who has been conducting related research on OAM.

### Disclosures

The authors declare no conflicts of interest.

### Data availability

Data underlying the results presented in this paper are not publicly available at this time but may be obtained from the authors upon reasonable request.

### Supplemental document

See Supplement 1 for supporting content.

# (Supplemental) Inverse-Designed Non-Hermitian Hollow Nanowire Cavity for Generating Optical Orbital Angular Momentum

XUEN ZHEN LIM,[1,†] MASATO TAKIGUCHI,[1,2,†,*] RYUJI KURUMA,[3] HISASHI SUMIKURA[1,2], AND MASAYA NOTOMI[1,2,3]

[1]Basic Research Laboratories, NTT Corp., 3-1, Morinosato Wakamiya, Atsugi, Kanagawa 243-0198, Japan
[2]NTT Nanophotonics Center, NTT Corp., 3-1, Morinosato Wakamiya, Atsugi, Kanagawa 243-0198, Japan
[3]Department of Physics, Institute of Science Tokyo, 2-12-1 Ookayama, Meguro-ku, Tokyo 152-8550, Japan
[†]These authors contributed equally to this work.
*masato.takiguchi@ntt.com

## SUPPLEMENTAL 1: FUZZY SELF-TUNED PARTICLE SWARM OPTIMIZATION

We employed the Fuzzy Self-Tuned Particle Swarm Optimization (FST-PSO) algorithm to address complex, high-dimensional optimization tasks in computational simulations. This method builds upon the classical Particle Swarm Optimization (PSO) framework, enhancing it through the integration of fuzzy logic to dynamically adjust the key hyperparameters governing particle behavior. In traditional PSO, particles explore a search space by updating their positions according to a linear combination of their velocity of previous timestep, displacement from current personal best-known positions and displacement from current global best across the swarm, weighted by inertia, cognitive, and social coefficients respectively. Additionally, the maximum (*L)* and minimum (*U)* velocities of particles can also be implemented to further control exploration behavior. While effective in many scenarios, classical PSO requires careful manual tuning of these coefficients to balance convergence speed against the risk of stagnation in local optima. This static approach often leads to premature convergence or inefficient exploration when dealing with nontrivial objective landscapes.

Figure S1 illustrates how particles converge toward a local minimum over multiple time steps. The velocity of the *n*-th particle can be expressed by the following equation.

$$v_n = w_{n-1} \cdot \boldsymbol{v_{n-1}} + c_{soc,n-1} \cdot \boldsymbol{R}_1 \cdot (x_{n-1} - g_{n-1}) + c_{cog,n-1} \cdot \boldsymbol{R}_2 \cdot (x_{n-1} - b_{n-1}).$$

Here, $w$, $c_{soc,n-1}$, $c_{soc,n-1}$ and $\boldsymbol{R}$ are an inertial factor, social factor, cognitive factor, and random vector. $x_{n-1} - g_{n-1}$ and $x_{n-1} - b_{n-1}$ represent displacement from global best pos. The maximum and minimum velocities of the *n*-th particle can be expressed as follows.

$$\begin{gathered} v_{min,n} = U_n(b_{max,n} - b_{min,n}) \\ v_{max,n} = L_n(b_{max,n} - b_{min,n}) \\ v_n \in [v_{min,n}, v_{max,n}] \\ \boldsymbol{x_n} = \boldsymbol{x_{n-1}} + \boldsymbol{v_n} \end{gathered}$$

Here $U_n$ and $L_n$ are minimum and maximum velocity factors. The velocity $v_n$ is constrained between the minimum and the maximum velocity. At the n-th time step, the particle's position is updated by adding its current velocity to the position from the previous step.

FST-PSO overcomes these limitations by replacing fixed hyperparameter settings with a fuzzy logic controller that responds in real time to the search process. The controller takes two inputs: $\boldsymbol{\phi}$, a normalized measure of fitness improvement between iterations, and $\boldsymbol{\delta}$, the positional distance between a particle's current location and its personal best. Using the respective membership functions (Fig. S2), the linguistic values of $\boldsymbol{\phi}$ and $\boldsymbol{\delta}$ are determined, which are then used with a fuzzy rule set to evaluate the output values of each hyperparameter. As a result, the particles are able to shift their behavior dynamically—favoring exploration when little progress is detected and reinforcing convergence when improvement is steady.

We implemented this optimization scheme within a parallel architecture that integrates seamlessly with COMSOL Multiphysics. Particle positions are initialized randomly within prescribed bounds and distributed across multiple COMSOL worker sessions. The framework supports multiprocessing, enabling simultaneous evaluation of fitness across all particles. To ensure robustness and recoverability, the system checkpoints particle positions, velocities, and internal states after each iteration. In the event of hardware interruption, license limitations, or intentional process suspension, the optimization can resume from the most recent saved state without loss of progress. Furthermore, when a particle is deemed to have stagnated—i.e., no longer contributing to the exploration of the solution space—it is reinitialized at a random position, thereby reintroducing diversity and reducing the likelihood of becoming trapped in suboptimal regions.

Compared to the commonly used Nelder-Mead (NM) method, FST-PSO demonstrated improved performance both in terms of convergence speed and final objective value. It consistently reached higher-quality solutions and exhibited more stable behavior across a range of test scenarios. A significant advantage of this approach is its elimination of manual hyperparameter tuning, allowing for a more automated and reproducible optimization pipeline. The combination of fuzzy self-tuning and inherent parallelism makes FST-PSO particularly well suited for computationally intensive applications where both robustness and efficiency are critical.

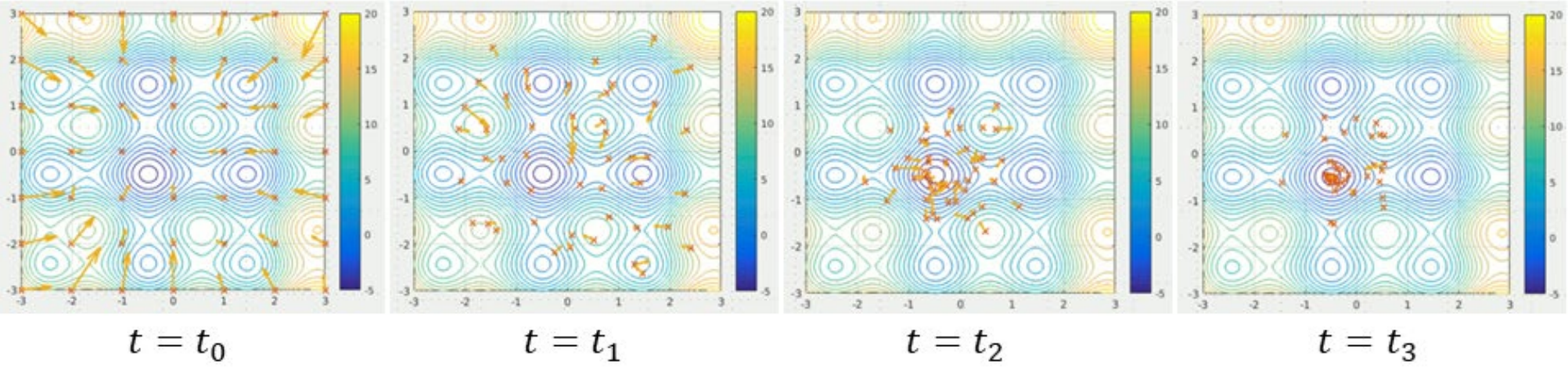

Fig. S1. Images depicting the convergence of particles across multiple time steps towards a local minimum.

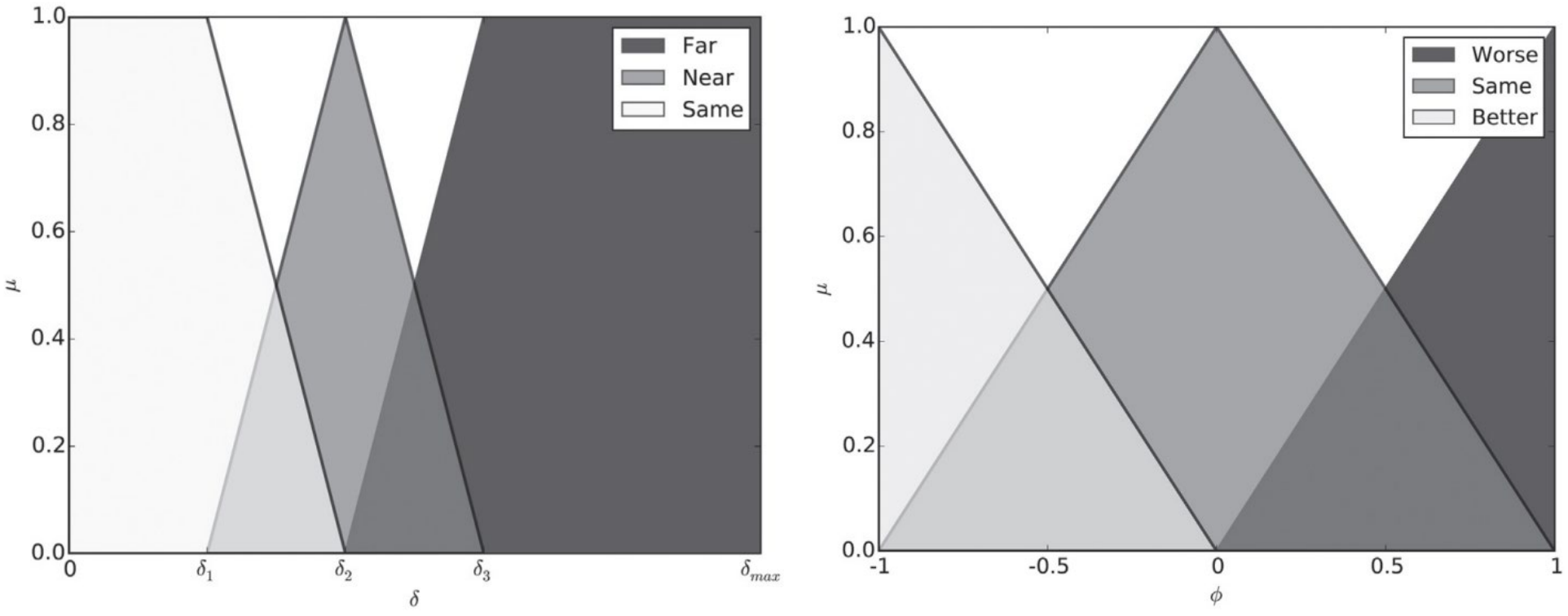

Fig. S2. Membership functions of the linguistic values of $\boldsymbol{\delta}$ (left) and $\boldsymbol{\phi}$ (right). For $\boldsymbol{\delta}$, the following heuristic is used: $\boldsymbol{\delta}_1 = 0.2 \cdot \boldsymbol{\delta}_{max}, \boldsymbol{\delta}_2 = 0.4 \cdot \boldsymbol{\delta}_{max}, \boldsymbol{\delta}_3 = 0.6 \cdot \boldsymbol{\delta}_{max}$.

| **Rule No.** | **Rule Definition** |
|---|---|
| 1 | if ($\boldsymbol{\phi}$ is *Worse* or $\boldsymbol{\delta}$ is *Same*) then (*Inertia* is *Low*) |
| 2 | if ($\boldsymbol{\phi}$ is *Same* or $\boldsymbol{\delta}$ is *Near*) then (*Inertia* is *Medium*) |
| 3 | if ($\boldsymbol{\phi}$ is *Better* or $\boldsymbol{\delta}$ is *Far*) then (*Inertia* is *High*) |
| 4 | if ($\boldsymbol{\phi}$ is *Better* or $\boldsymbol{\delta}$ is *Near*) then (*Social* is *Low*) |
| 5 | if ($\boldsymbol{\phi}$ is *Same* or $\boldsymbol{\delta}$ is *Same*) then (*Social* is *Medium*) |
| 6 | if ($\boldsymbol{\phi}$ is *Worse* or $\boldsymbol{\delta}$ is *Far*) then (*Social* is *High*) |
| 7 | if ($\boldsymbol{\delta}$ is *Far*) then (*Cognitive* is *Low*) |
| 8 | if ($\boldsymbol{\phi}$ is *Worse or Same,* or $\boldsymbol{\delta}$ is *Same or Near*) then (*Cognitive* is *Medium*) |
| 9 | if ($\boldsymbol{\phi}$ is *Better*) then (*Cognitive* is *High*) |
| 10 | if ($\boldsymbol{\phi}$ is *Same or Better,* or $\boldsymbol{\delta}$ is *Far*) then (*L* is *Low*) |
| 11 | if ($\boldsymbol{\phi}$ is *Same* or $\boldsymbol{\delta}$ is *Near*) then (*L* is *Medium*) |
| 12 | if ($\boldsymbol{\phi}$ is *Worse*) then (*L* is *High*) |
| 13 | if ($\boldsymbol{\delta}$ is *Same*) then (*U* is *Low*) |
| 14 | if ($\boldsymbol{\phi}$ is *Same or Better,* or $\boldsymbol{\delta}$ is *Near*) then (*U* is *Medium*) |
| 15 | if ($\boldsymbol{\phi}$ is *Worse* or $\boldsymbol{\delta}$ is *Far*) then (*U* is *High*) |

Table. S1. Table containing set of Fuzzy rules matching the indicators' linguistic values to the appropriate terms for the output hyperparameter variables

| **Variables** | **Term** | | |
|---|---|---|---|
| | **Low** | **Medium** | **High** |
| *Inertia* | 0.3 | 0.5 | 1.0 |
| *Social* | 1.0 | 2.0 | 3.0 |
| *Cognitive* | 0.1 | 1.5 | 3.0 |
| *L* | 0.0 | 0.001 | 0.01 |
| *U* | 0.1 | 0.15 | 0.2 |

Table. S2. Table containing output hyperparameter variables and their respective defuzzification

## SUPPLEMENTAL 2: COMPARISON BETWEEN NORME AND POYNTING VECTOR

Anti-nodes in normE becomes more 'smeared' as we capture the time average of a traveling topological light (Fig. S3), as well as strong chirality in Poynting vector distribution.

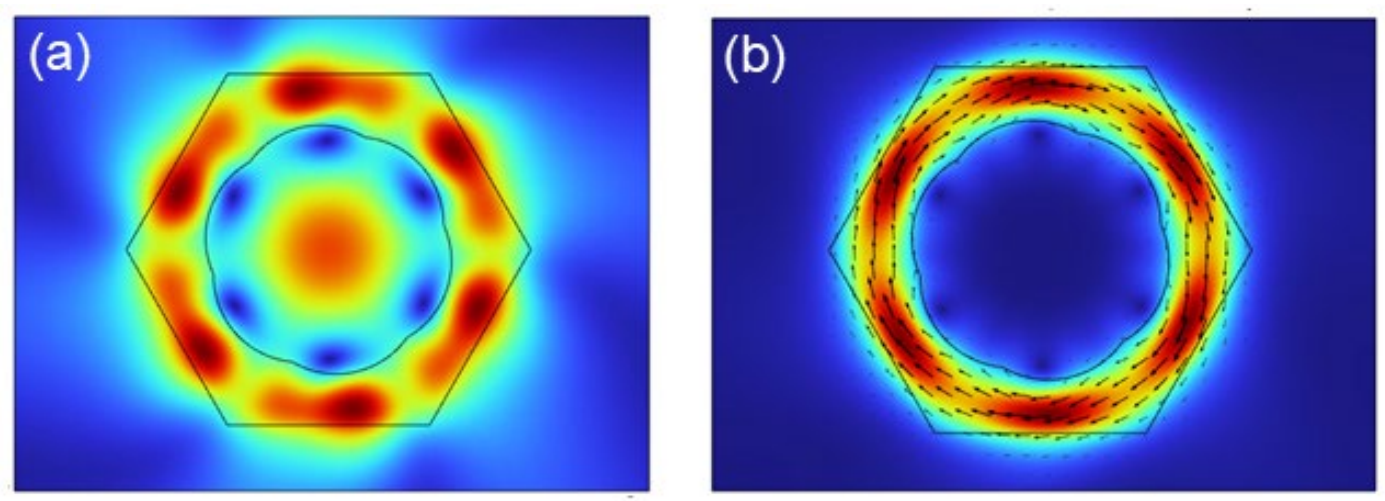

Fig. S3. Field distribution of optimized nanowire design, expressed by: (a) normE (b) Poynting vector

In comparison, we observe isolated anti-nodes in normE of even and odd parity near-degenerate modes in a regular GaN nanowire with circular hollow core, indicating the lack of chiral movement (Fig. S4). This is supported by the lack of curl in the Poynting vector distribution of the respective modes.

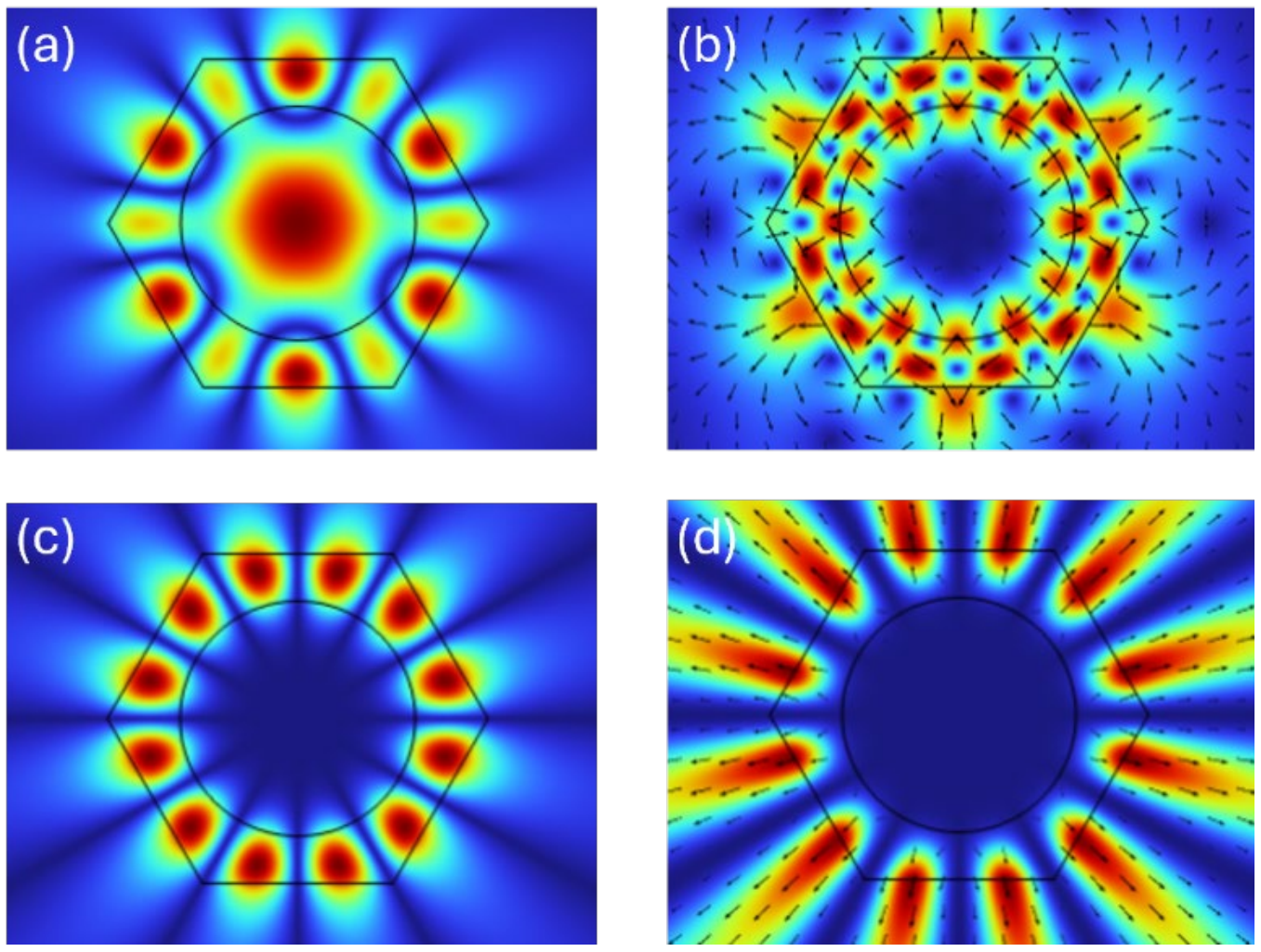

Fig. S4. Field distribution of unoptimized circular hollow nanowire design, expressed by: (a) normE of even parity mode (b) Poynting vector of even parity mode (c) normE of odd parity mode (d) Poynting vector of odd parity mode

## Supplemental 3: Width Profile of nanowire

Structural asymmetry plays a crucial role in the generation of OAM. Here, we evaluated the GaN thickness when rotating the central airhole cluster of a hexagonal GaN nanowire. As shown in the Fig S4, the structure retains rotational symmetry when the center consists of a simple single airhole or when $\Delta\theta = 0$. However, at $\Delta\theta = 12$, the structure becomes asymmetric with respect to the rotation direction.

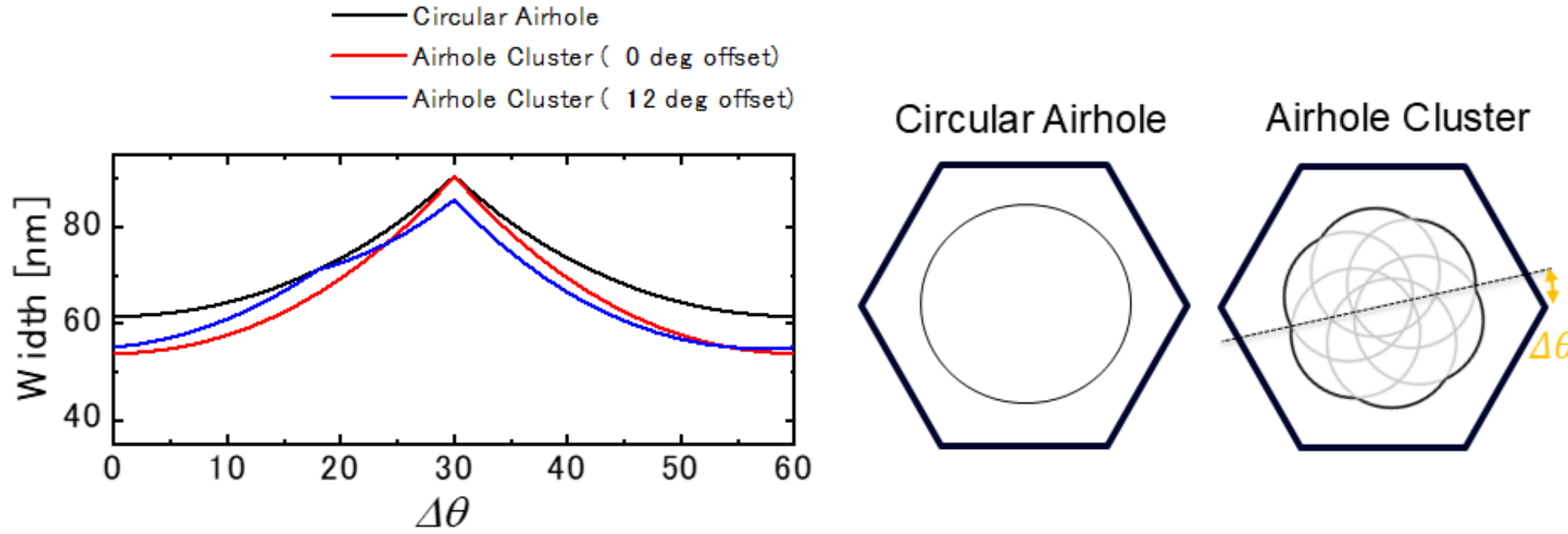

Fig. S5. Width profile of hexagonal nanowire design. The black line shows the width profile of a regular hexagonal nanowire with circular central airhole. The red and blue lines show the width profile of the hexagonal nanowire with optimized airhole cluster for 0- and 12-degree offsets.

## SUPPLEMENTAL 4: TWO MODE APPROXIMATION

For mediums supporting topological eigenmodes of azimuthal order |m|, like WGMs, there exists a clockwise and counterclockwise (+m and -m) traveling wave basis, which can also be represented with a pair of standing wave basis (even and odd parity modes). Like spin angular momentum with linear and circular polarizations, the two sets of bases of WGMs can be mapped onto a Poincare sphere (Fig. S6), with +m and -m on the poles, and the even and odd parity modes along the equator ($\theta = \frac{\pi}{2}$), separated by a relative phase of $\varphi = \pi$. A general equation for an eigenstate on the Poincare sphere can be expressed as:

$$|\boldsymbol{\psi}\rangle = cos(\theta)| + m\rangle + e^{i\varphi}\, sin(\theta)| - m\rangle$$

Fig. S6. Poincare sphere with the traveling and standing wave bases mapped onto.

The two-mode approximation method is commonly used to estimate the non-Hermitian effects arising from perturbation(s) introduced to Hermitian Hamiltonian systems. We first take the standing wave Hamiltonian of the unperturbed system. For WGMs, it is usually a mirror symmetric, circular resonator at a diabolic point (degenerate eigenfrequencies and independent eigenstates). To account for the effects of a perturbation on the unperturbed Hamiltonian $H_0$, we treat the introduction of a perturbation source as a Hamiltonian $H_1$:

$$H_0 = \begin{pmatrix} \Omega_0 & 0 \\ 0 & \Omega_0 \end{pmatrix}, H_1 = \begin{pmatrix} 2V_1 & 0 \\ 0 & 2U_1 \end{pmatrix}$$

Where $\Omega_0$is the unperturbed degenerate frequency; with $2V_1$and $2U_1$ being the frequency shifts in the standing even and odd parity modes as a result of perturbation introduction, which can be computed in simulation (FEM or FDTD). In Wiersig's case, it is from size-variant nanoparticles positioned around the edges of a disk. To create an analogue to Wiersig's model in the GaN nanowire system, we treat the nanowire cross section as a ring resonator-like cavity, with deformations added to the inner and outer edges, formed by the airhole cluster and the hexagonal outer boundary respectively. The deformations are treated as an equivalent to the nanoparticles in Wiersig's, for in both cases they act as scatterers that alter the eigenfrequencies of the near-degenerate parity mode, break the mirror symmetry of the system, and induced non-Hermicity (when positioned optimally in the azimuthal direction). We then determine the unperturbed Hamiltonian $H_0$ and perturbation Hamiltonians for the inner and outer deformations respectively.

The non diagonal elements represent the coupling shifts between the parity mode pairs but are negligible in value when examining each scatterer individually (also true in Wiersig's case). We then follow this up by transforming our standing wave basis Hamiltonians $H$ to traveling wave basis Hamiltonians $\widetilde{H}$ using the transformation matrix $M^{\dagger}$ (note that $H_0$ transformation is trivial):

$$M^{\dagger} = \begin{pmatrix} 1/\sqrt{2} & -i/\sqrt{2} \\ 1/\sqrt{2} & i/\sqrt{2} \end{pmatrix}, \widetilde{H_1} = M^{\dagger} H_1 M$$

In Wiersig's, all perturbation Hamiltonians (each representing a particle) are then added to the unperturbed Hamiltonian (in traveling wave basis). For each particle, the non-diagonal elements are multiplied by a phase shift factor of $e^{-i2m\beta_j}$, where $\beta_j$ is the angular offset (radians) from a designated origin angle and m is the azimuthal order of WGM. The same is performed for the GaN nanowire model, but with the relative angle between the inner and outer perturbations.

$$\widetilde{H} = \widetilde{H_0} + \sum_{j=1}^{N} \widetilde{H_j} = \begin{pmatrix} \Omega_0 & 0 \\ 0 & \Omega_0 \end{pmatrix} + \sum_{j=1}^{N} \begin{pmatrix} V_j + U_j & [V_j - U_j]e^{-i2m\beta_j} \\ [V_j - U_j]e^{i2m\beta_j} & V_j + U_j \end{pmatrix} = \begin{pmatrix} \Omega & A \\ B & \Omega \end{pmatrix}$$

From the resulting Hamiltonian, we compute the predicted standing wave eigenfrequencies using $\Omega\pm = \pm\sqrt{AB}$. The mode overlap $S_T$ between the WGM eigenmode pair is the normalized difference between the non-diagonal Hamiltonian components (traveling wave basis scattering coefficients) and can subsequently be used to compute chirality $\alpha$.

$$S_T = \left| \frac{|A| - |B|}{|A| + |B|} \right|, \alpha = \frac{2S_T}{1 + S_T}$$

The key values for the GaN nanowire system are as follows, note that there is a set of frequency shifts for inner perturbation and one for outer perturbation:

| Variables | Values (Hz) |
|---|---|
| $\Omega_0$ | 8.4303E14+6.4426E11i |
| $2V_{outer}$ | -7.7900E12 + 1.7226E12i |
| $2U_{outer}$ | -5.3400E12 + 4.9460E10i |
| $2V_{inner}$ | -6.1000E11 + 2.1090E10i |
| $2U_{inner}$ | -4.8000E+11 - 3.2300E09i |

Table. S3. Variable values used for two mode approximation

The mode overlap is then calculated over a range of angular offsets and compared to simulated outcomes. However, sweeping across $\Delta\theta$, we observed that not only does the prediction underestimates mode overlap, the period of phase change also appears to be different (Fig. S7). In Wiersig's, the model suggests a period of $\frac{\pi}{m}$ when sweeping across the angular offset between perturbation source, while our finding points to a period of $\frac{2\pi}{m}$. Indeed, Wiersig has expressed that the model functions best when nanoparticles are smaller than the anti-nodes of the WGM field distribution and strictly act as point-like dipole scatterers ($r \ll \lambda$). It is evident that the nanoparticle model is not directly compatible with ours, likely due to our relatively larger perturbation structures with more complex interactions between the many components. For instance, the GaN nanowire's outer perturbation structure spans across multiple anti-nodes (Fig. S8). In Fig. S9, we iteratively remove one perturbation structure from the inner edge to observe their overall contribution to OAM generation (accounting for all potential permutations). We found that the count of inner perturbation almost linearly correlates with OAM generation. We repeat this procedure but with only outer perturbation structure instead (Fig. S10), and it unveiled an irregular, non-linear contribution to OAM, suggesting that the large outer structures do possess complicated non-Hermitian effects. Additionally, the interaction between singular inner and outer perturbative structure around a ring resonator shows a chaotic correlation to mode overlap (Fig. S11). Nevertheless, we believe it can still be reasonably implied that both Wiersig's and this GaN nanowire models likely function similarly at a qualitative level: non-Hermitian systems enabled through angular offsets of varied perturbation structures. Future investigations would be needed to determine a quantitative two-mode approximation model that generalize to larger perturbation structures with accuracy, especially one that accounts for perturbation effects coming from both the inner and outer edges of a ring resonator.

It is possible that we can modify the geometry of the outer perturbative structures to induce a simpler, more predictable perturbation model. That said, this model's benefit lies in its fabrication simplicity, whereby we leverage the inherent hexagonal cross-section of GaN nanowires (arising from GaN's lattice structure) as a source of perturbation and supplement it with minimal modifications to create a near-EP non-Hermitian system. Potential future research could involve exploring other variants of inner and outer perturbation geometries applied to other ring resonator materials at different scales.

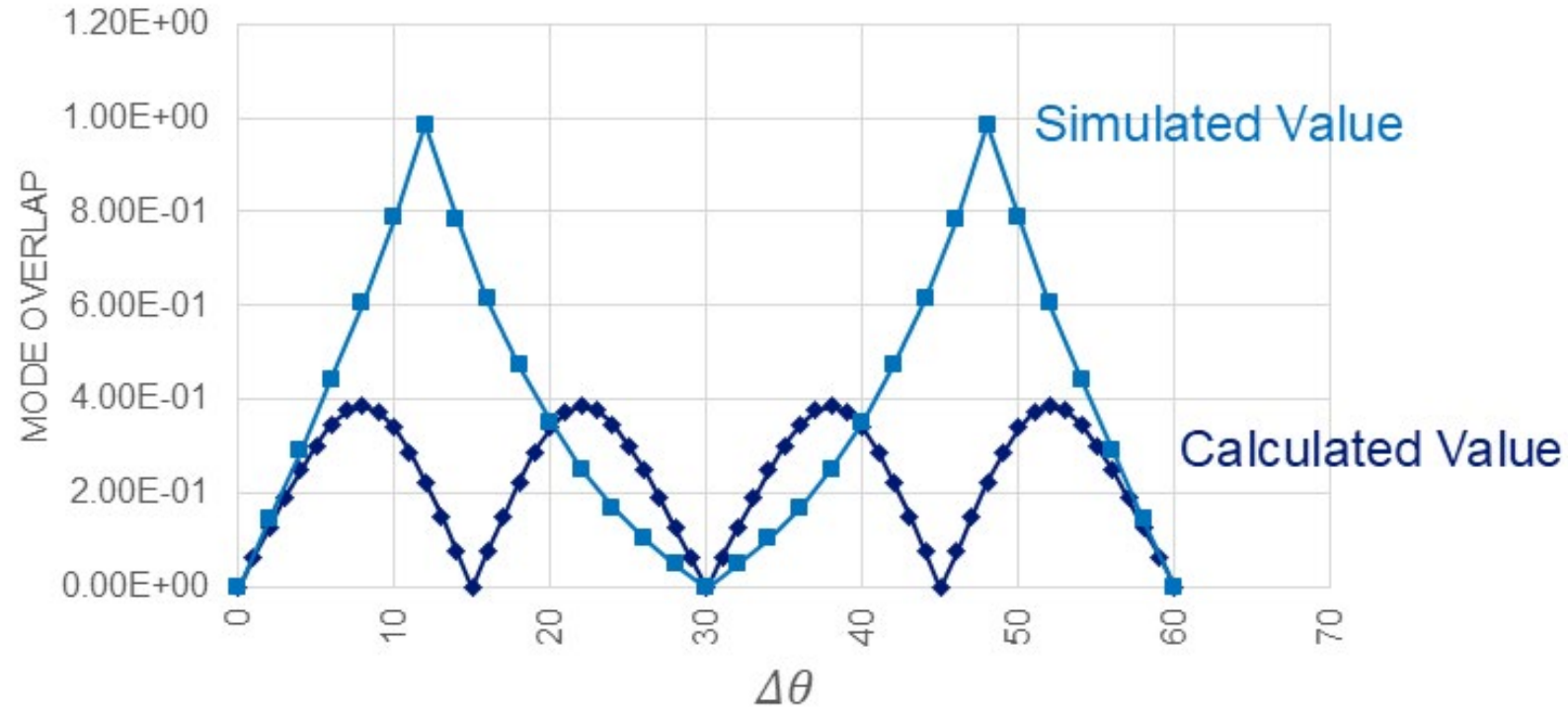

Fig. S7. Comparison between simulated and calculated overlap for GaN nanowire model. The calculated model using two-mode approximation predicts weaker overlap and also suggests a 30° period, which corresponds to $\frac{\pi}{m} rad$ where m is 6. The simulated value instead demonstrates a period of $\frac{2\pi}{m} rad$.

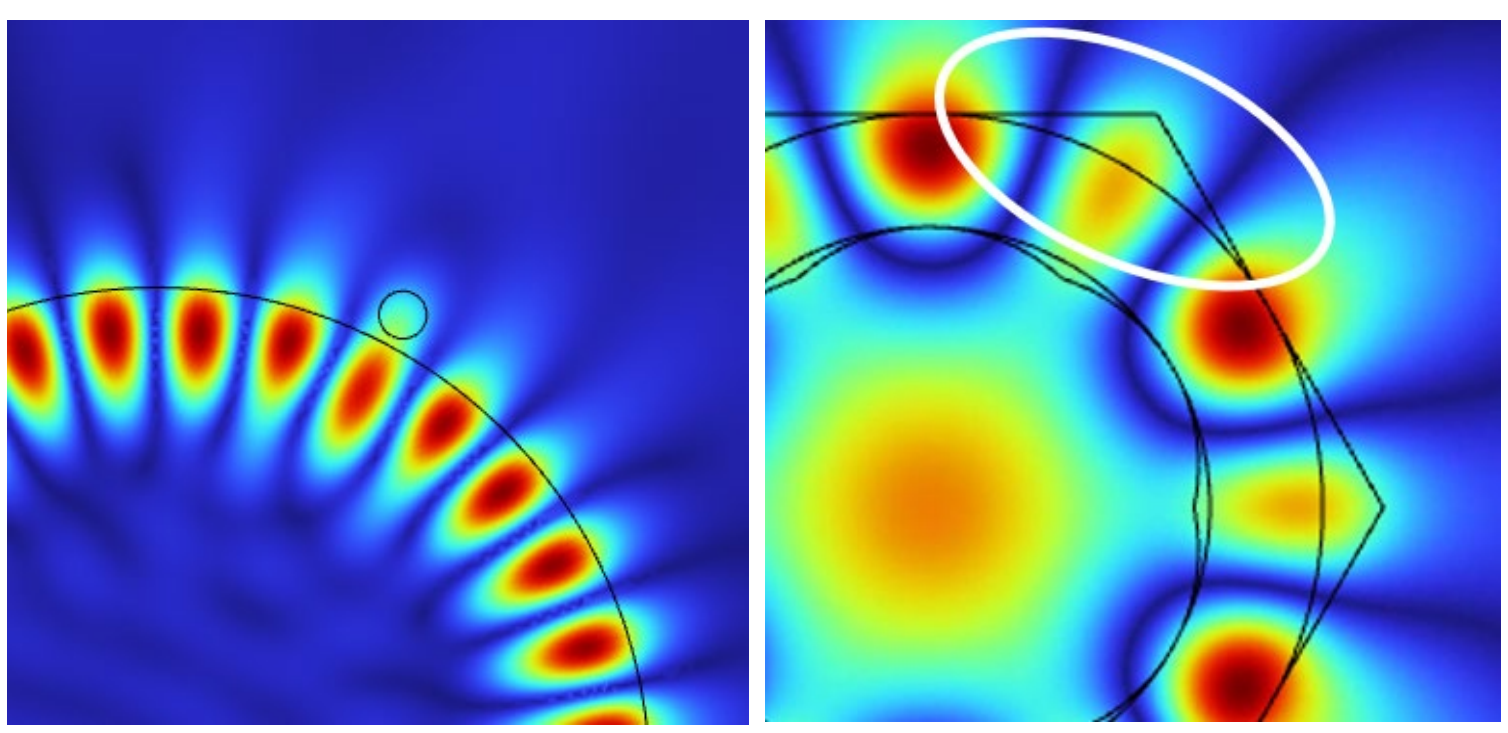

Fig. S8. Left: Wiersig's model showing perturbative nanoparticle interacting with a single anti-node. Right: GaN nanowire model, where outer perturbative deformation spans across multiple anti-nodes.

| θ | Variable | Calculated Value | Simulated Value | %Err |
|---|---|---|---|---|
| 0° | Ω − | 7.9341E14 + 1.096E13i | 8.0983E14+1.7598E12i | (2–80i)% |
| | Ω + | 8.0733E14 + 1.07E12i | 8.1141E14+6.1809E11i | (0.5-40i)% |
| | S | 0 | 2.16E-4 | ~0% |
| | α | 0 | 4.33E-4 | ~0% |
| 12° | Ω − | 7.9270E14 + 1.1109E13i | 8.1088E14+1.2050E12i | (2-89i)% |
| | Ω + | 8.0804E14 + 9.4E11i | 8.1110E14+1.3137E12i | (0.2-40i)% |
| | S | 0.18888 | 0.98043 | 419% |
| | α | 0.31775 | 0.99012 | 212% |

Table. S4. Comparing calculated and simulated values at diabolic ($\Delta\theta$=0°).) and exceptional points ($\Delta\theta$=12°).

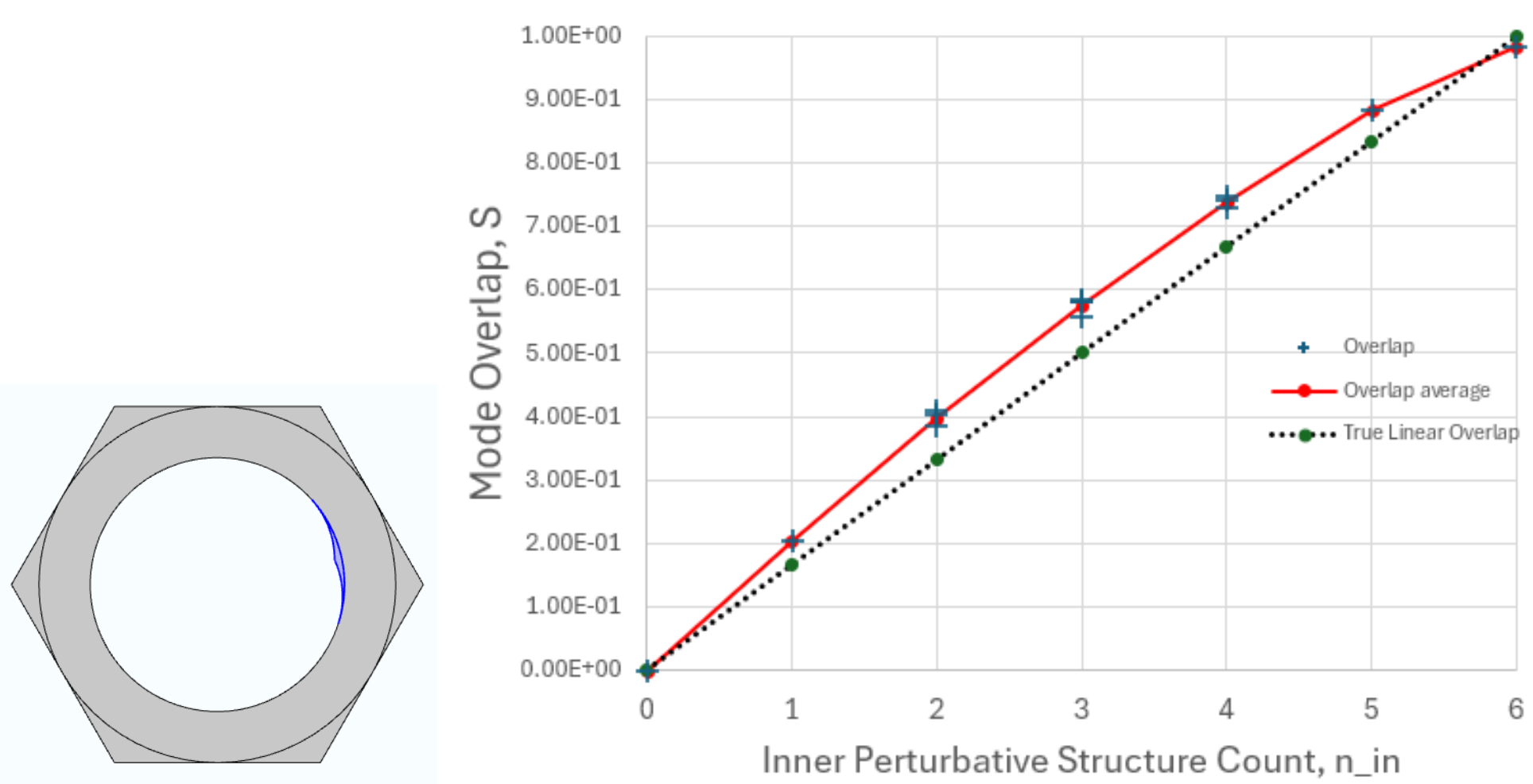

Fig. S9. Left: Image showing an example of all six outer perturbation present but only a single inner perturbative structure (highlighted), corresponds to $n_{in}$=1 on the plot. Right: Plot shows the correlation between the number of inner perturbative structure vs mode overlap. Cross marker indicates the mode overlap calculated for all the potential circular permutation for a particular number of perturbative structures. Results show mode overlaps are similar for any count regardless of permutation.

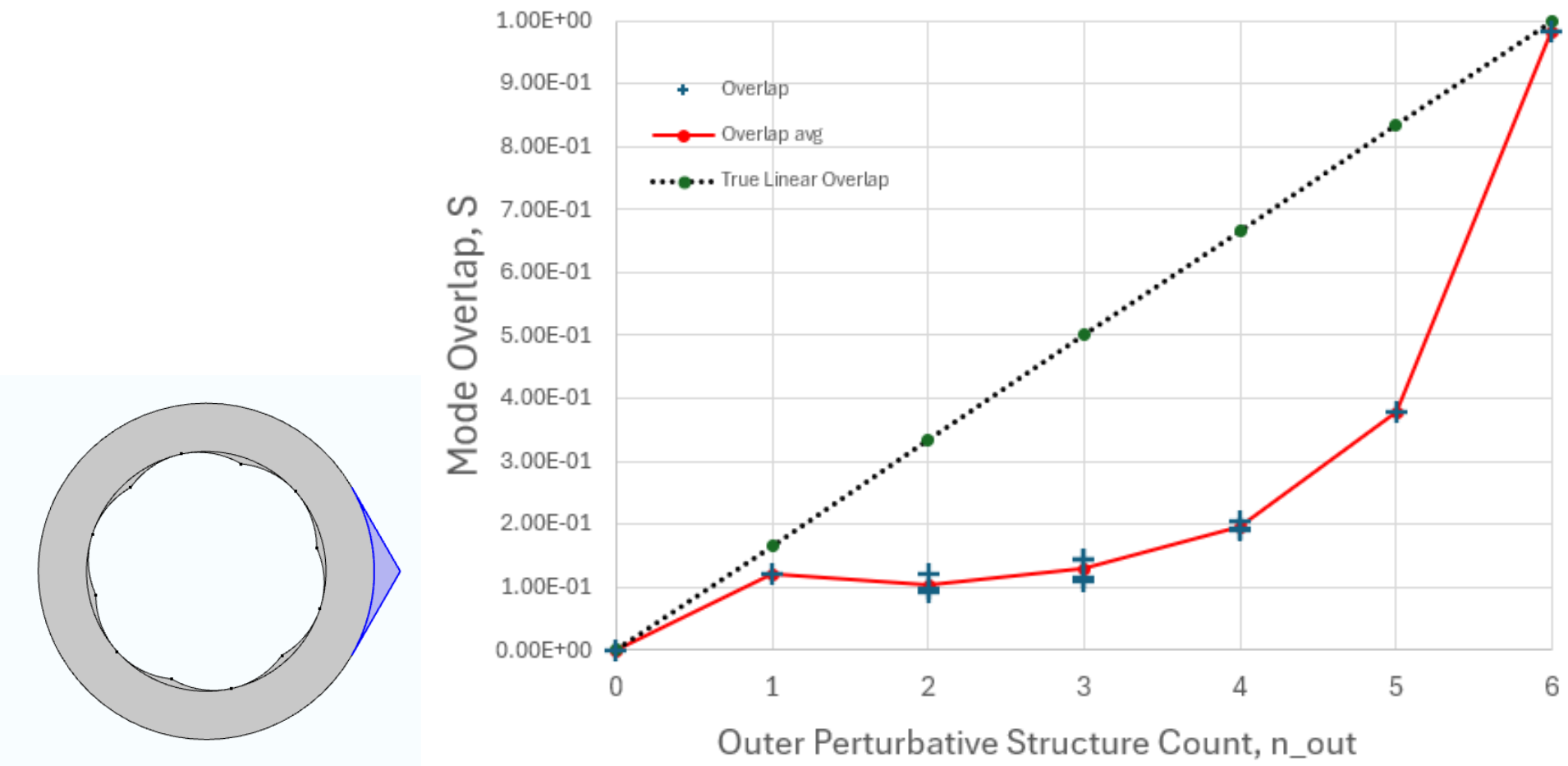

Fig. S10. Left: Image showing an example of all six inner perturbation present but only a single outer perturbative structure (highlighted), corresponds to $n_{out}$=1 on the plot. Right: Plot shows the correlation between the number of outer perturbative structure vs mode overlap (right). Cross marker indicates the mode overlap calculated for all the potential circular permutation for a particular number of perturbative structures. Results show mode overlaps are similar for any count regardless of permutation.

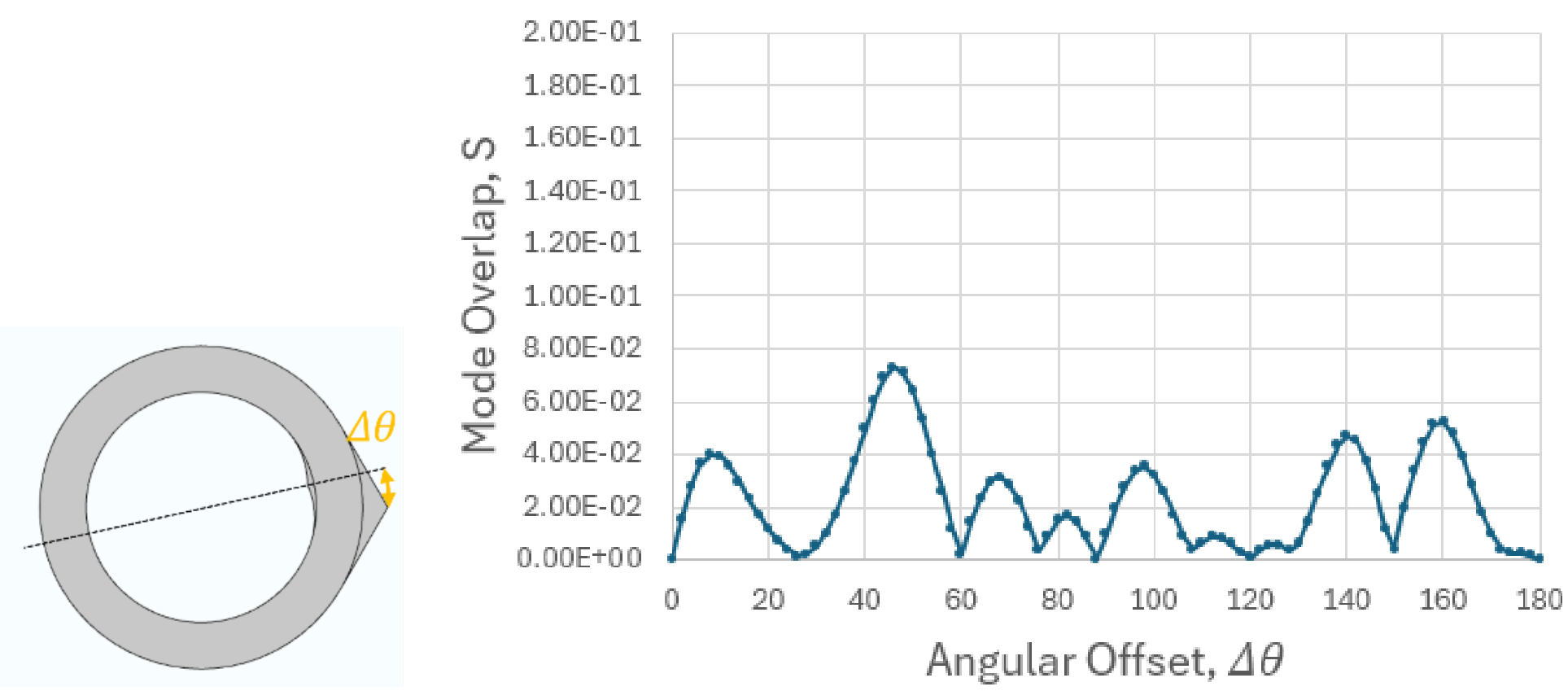

Fig. S11. Left: Image showing only one inner and one outer perturbation structure. Right: Plot showing the angular offset against mode overlap. Angles from 0° to 180° are checked as geometries from 180° to 360° are symmetrical.

## Supplemental 5: Model for Fabrication Errors

To introduce fluctuations in the cross-sectional geometry, we constructed the hexagon and air-hole cluster using polar equations and the Parametric Curve feature in COMSOL Multiphysics. The hexagonal outer nanowire boundary is defined by Equation (5.1) and using the hexagonal radius $r_{hex}$ defined previously.

$$\rho_{hex}(\theta) = \frac{\frac{\sqrt{3}}{2} r_{hex}}{cos(mod\left(\theta, \frac{\pi}{3}\right) - \frac{\pi}{6})} \qquad (5.1)$$

Next, the central airhole cluster is constructed by first defining a circle with radius $r_{hole}$, shifted in the y-axis by constant R, and then shifted azimuthally by $\Delta\theta$, as shown in Equation 5.2. The full cluster was then built using Equation 5.3, which finds the maximum of six azimuthally distributed circles from Equation 5.2.

$$\rho_{circle}(\theta, \phi) = R \cdot sin(\theta + \phi + \Delta\theta) \pm \sqrt{r_{hole}{}^2 - R^2 cos^2(\theta + \phi + \Delta\theta)} \qquad (5.2)$$

$$\rho_{cluster}(\theta) = max_{k=0}^{5} \left[\rho_{circle}\left(\theta, \frac{k\pi}{3}\right)\right] \qquad (5.3)$$

We defined a simple sinusoidal equation (Equation 8) to approximate roughness at the inner and outer surface of the nanowire, controlled by amplitude $A$ and an angular wavenumber $B$[1/rad]. Equation 5.4 was then added to Equation 5.1 and 5.3.

$$\rho_{defect}(\theta) = A \cdot sin(B\theta) \qquad (5.4)$$

## SUPPLEMENTAL 6: MESH CONFIGURATION (3D)

We devised a method that imposes symmetry on the mesh structure independently of the actual physical symmetry of the device. Specifically, for asymmetric geometries such as air hole arrays embedded in GaN substrates, we extracted and mirrored mesh boundary outlines to artificially restore a form of geometric balance (Fig. S12). This mesh-symmetry enforcement serves solely as a meshing guide and does not alter the actual physical structure or material properties. In practice, this allows consistent meshing patterns across replicated domains, which in turn significantly reduces computational noise and artificial OAM contributions.

An important consideration in our implementation is the limitation of commercial FEM software such as COMSOL, which only permits mesh replication under exact mirror symmetry conditions. Thus, careful design of the meshing template and boundaries was essential to exploit COMSOL's mesh-copying capabilities without compromising the physical accuracy of the model.

To further optimize computational efficiency, we employed swept meshing in regions where geometry allowed it. Compared to conventional tetrahedral meshing, swept meshes achieved equivalent numerical accuracy while requiring only half the degrees of freedom (DOF). In prism-shaped elements, this reduction was even more pronounced, using as little as 1/7 the number of elements. This significantly decreased simulation time and memory consumption.

We also incorporated boundary layer meshing near critical regions such as high-index contrast interfaces and scattering boundaries. By refining mesh fidelity in the direction of wave propagation and maintaining coarser meshing in distant regions, we preserved the accuracy of absorbing boundary conditions (e.g., SBC surface) without sacrificing performance. This approach allows us to reduce the meshing fidelity of the space away from the nanowire structure itself but reintroduce high resolution meshing as it approaches the boundaries.

Benchmark simulations were conducted using a rotationally symmetric structure where the expected OAM was analytically known to be zero. When using our symmetric meshing approach, the simulation correctly reproduced this null-OAM condition, validating the mesh strategy. In contrast, meshing schemes that did not respect symmetry often exhibited artificial OAM behavior, underscoring the importance of consistent meshing in such systems.

Overall, our meshing methodology not only ensures reliable simulation of OAM phenomena in asymmetric structures, but also provides significant reductions in computational cost. These advantages make it a valuable approach for large-scale 3D simulations of integrated photonic devices where both precision and efficiency are critical.

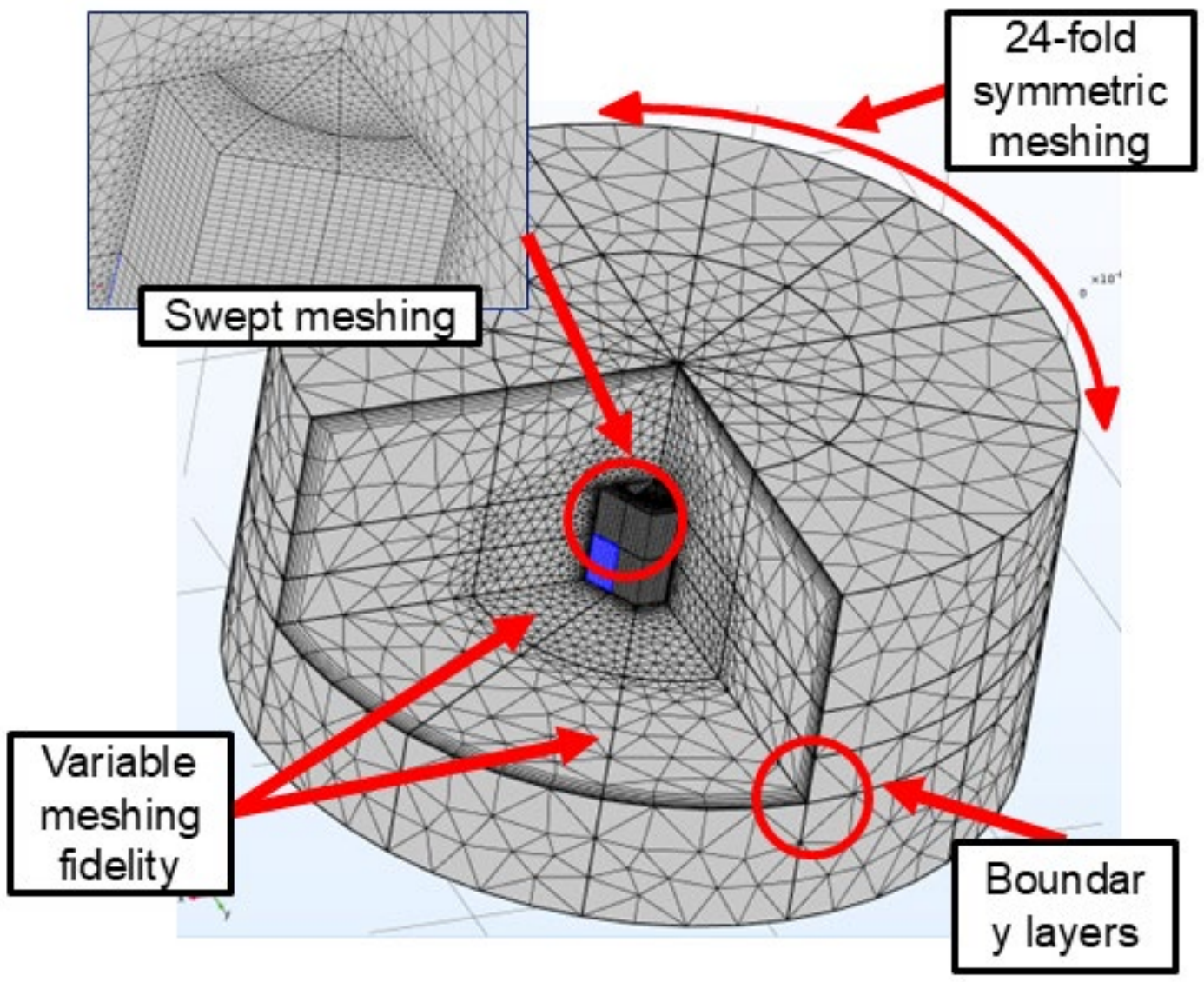

Fig. S12. Mesh structure used in the 3D calculation. A 24-fold symmetric meshing scheme is applied to reduce the computational domain. Swept meshing is used along the axial direction, while variable mesh density is introduced to refine the mesh near the nanowire and other regions of interest. Boundary layers are also included to accurately resolve the fields near the surfaces.